


%
%

%

\font\Bigrm=cmr10 scaled \magstep3

\font\bigbf=cmssbx10 scaled \magstep2

%

%



\font\eightrm=cmr8
\font\sixrm=cmr6
\font\fiverm=cmr5
\font\eighti=cmmi8
\font\sixi=cmmi6
\font\fivei=cmmi5
\font\eightsy=cmsy8
\font\sixsy= cmsy6
\font\fivesy=cmsy5
\font\tenex=cmex10
\font\eightit=cmti8
\font\eightsl=cmsl8
\font\eighttt=cmtt8
\font\eightbf=cmbx8

\font\sixbf=cmbx6
\font\fivebf=cmbx5
\newskip\ttglue
\def\eightpoint{\def\rm{\fam0\eightrm}
 \textfont0=\eightrm \scriptfont0=\sixrm \scriptscriptfont0=\fiverm
 \textfont1=\eighti  \scriptfont1=\sixi  \scriptscriptfont1=\fivei
 \textfont2=\eightsy \scriptfont2=\sixsy \scriptscriptfont2=\fivesy
 \textfont3=\tenex   \scriptfont3=\tenex \scriptscriptfont3=\tenex
 \textfont\itfam=\eightit \def\it{\fam\itfam\eightit}
 \textfont\slfam=\eightsl \def\sl{\fam\slfam\eightsl}
 \textfont\ttfam=\eighttt \def\tt{\fam\ttfam\eighttt}
 \textfont\bffam=\eightbf \scriptfont\bffam=\sixbf
  \scriptscriptfont\bffam=\fivebf \def\bf{\fam\bffam\eightbf}
   \tt \ttglue=.5em plus.25em minus.15em
 \normalbaselineskip=10pt
 \setbox\strutbox=\hbox{\vrule height7pt depth2pt width0pt}
 \let\sc=\sixrm \let\big=\eightbig \normalbaselines\rm}
\def\tenbig#1{{\hbox{$\left#1\vbox to8.5pt{}\right.\n@space$}}}

\def\cosTD{\mathop{\rm cos3D}\nolimits}

\def\bR{{\Bbb R}}
\def\bS{{\Bbb S}}
\def\bT{{\Bbb T}}

\def\bZ{{\Bbb Z}}

\chardef\o="1C
\def\frac#1#2{{#1\over #2}}

\def\leaderfill{\leaders\hbox to 1em{\hss.\hss}\hfill}
\def\tZ{\bZ^3}
\def\dS{\bS^2}
\def\dT{\bT^2}
\def\tR{\bR^3}

\def\tT{\bT^3}

\def\rmU{\uppercase\expandafter{\romannumeral1}}
\def\rmD{\uppercase\expandafter{\romannumeral2}}
\def\rmT{\uppercase\expandafter{\romannumeral3}}

\newcount\chapnum                                                          
\newcount\firstpg
\newcount\refnumber                                                         
\newcount\secnum
\newcount\thmnum
\newcount\propnum
\newcount\defnum
\newcount\exnum
\newcount\eqnum
\newcount\subsecnum
\newcount\remnum
\newcount\picnum
\newcount\tablenum
\global\picnum=0
\global\tablenum=0
\global\remnum=0
\global\eqnum=0
\global\thmnum=0
\global\propnum=0
\global\defnum=0
\global\exnum=0
\global\secnum=0
\global\chapnum=0
\global\firstpg=1
\global\refnumber=0
\global\subsecnum=0
\def\currentsubsec{}

\def\makeheadline{\vbox to 0pt{\vskip-34.5pt                                 
    \line{\vbox to8.5pt{}\the\headline}\vss}\nointerlineskip}                

\def\title#1{\centerline{\Bigrm #1}\vskip 1truecm\def\lhead{}}
\def\author#1#2#3#4{
	\vskip 2.truecm
	\centerline{\bf #1}
	\centerline{\bf #2}
	\vskip .5truecm
	\centerline{#3}
	\centerline{#4}
}
\def\section#1#2{\global\advance\secnum by 1
    \bigskip\noindent
    {\bf\the\secnum\ \ #2}
    \medskip
    \def\sectno{\the\secnum}
    \mark{\the\secnum\enskip#2}
    \xdef\currentsec{\the\secnum}
    \xdef#1{\currentsec}
    \global\picnum=0
    \global\eqnum=0
    \global\remnum=0
    \global\thmnum=0
    \global\propnum=0
    \global\defnum=0
    \global\exnum=0     
    \global\subsecnum=0
}
\def\subsection#1{\global\advance\subsecnum by 1
    \medskip
    \line{\bf\ \the\secnum.\the\subsecnum\ \ #1\hfill}
    \medskip
    \def\subsectno{\the\secnum\the\subsecnum}
    \mark{\the\secnum\the\subsecnum\enskip#1}
    \xdef\currentsubsec{.\the\subsecnum}
    \global\picnum=0
    \global\eqnum=0
    \global\remnum=0
    \global\thmnum=0
    \global\propnum=0
    \global\defnum=0
    \global\exnum=0     
}
          
\long\def\prop#1#2{\global\advance\propnum by 1
        \xdef#1{Proposition \currentsec.\the\propnum}
        \bigbreak\noindent{\bf Proposition \currentsec.\the\propnum.}
        {\it#2} }
\long\def\define#1#2#3{\global\advance\defnum by 1
        \xdef#2{Definition \currentsec\currentsubsec.\the\defnum\ }
        \medbreak\noindent{\bf Definition \currentsec\currentsubsec.\the\defnum.}
        {\bf #1}{\sl#3} \medbreak}
\long\def\example#1#2{\global\advance\exnum by 1
        \xdef#1{Example \currentsec\currentsubsec.\the\exnum}
        \medbreak\noindent{\bf Example \currentsec\currentsubsec.\the\exnum.}
        {\sl#2} \medbreak}
\long\def\lemma#1#2{\global\advance\thmnum by 1
        \xdef#1{Lemma \currentsec\currentsubsec.\the\thmnum}
        \bigbreak\noindent{\bf Lemma \currentsec\currentsubsec.\the\thmnum.}
        {\sl#2}}
\long\def\rem#1#2{\global\advance\remnum by 1
        \xdef#1{Remark \currentsec\currentsubsec.\the\remnum\ }
        \bigbreak\noindent{\bf Remark \currentsec\currentsubsec.\the\remnum.}
        {\sl#2} }
\long\def\thm#1#2#3{\global\advance\thmnum by 1
        \xdef#2{Theorem \currentsec\currentsubsec.\the\thmnum\ }
        \medbreak\noindent{\bf Theorem \currentsec\currentsubsec.\the\thmnum.}
       {\bf #1}{\sl#3} \medbreak}
\long\def\cor#1#2{\global\advance\thmnum by 1
        \xdef#1{Corollary \currentsec\currentsubsec.\the\thmnum}
        \bigbreak\noindent{\bf Corollary \currentsec\currentsubsec.\the\thmnum.}
        {\sl#2} }
\long\def\conj#1#2{\global\advance\thmnum by 1
        \xdef#1{Conjecture \currentsec\currentsubsec.\the\thmnum}
        \bigbreak\noindent{\bf Conjecture \currentsec\currentsubsec.\the\thmnum.}
        {\sl#2} }

\long\def\caption#1#2{\vbox{\hsize 14.cm\eightbf Figure \picRef#1:\eightrm\ #2}}
\long\def\tableCaption#1#2{\vbox{\hsize 14.cm\noindent\eightbf Table \tableRef#1:\eightrm\ #2}}

\def\num{\global\advance\eqnum by 1
        \eqno({\rm\currentsec}.\the\eqnum)}
\def\picRef#1{\global\advance\picnum by 1
	\xdef#1{\currentsec.\the\picnum\ }
        \currentsec.\the\picnum\ }
\def\tableNum{\global\advance\tablenum by 1
        \eqno(\the\tablenum)}
\def\tableRef#1{\global\advance\tablenum by 1
	\xdef#1{\uppercase\expandafter{\romannumeral\the\tablenum}\ }
        \uppercase\expandafter{\romannumeral\the\tablenum}\ }
\def\eqalignnum{\global\advance\eqnum by 1
        ({\rm\currentsec}.\the\eqnum)}
\def\ref#1{\num  \xdef#1{(\currentsec.\the\eqnum)}}
\def\eqalignref#1{\eqalignnum  \xdef#1{(\currentsec.\the\eqnum)}}

\def\Acknowledgments{
	\vskip 1.cm{\bf Acknowledgments}\medskip
	\def\lhead{}
	\def\rhead{}
	\mark{}
}
\def\theBibliography{
	\vskip 2.truecm {\bigbf Bibliography} \vskip 1.truecm
	\def\lhead{}
	\def\rhead{}
	\mark{}
}

\def\bibitem{\vskip.1cm\par\noindent}

\def\vspace#1{\vcenter{\hbox{\hskip 1pt}\vskip #1}}                       
\headline={
\ifnum\pageno=\firstpg             
           \else                              
                 \ifodd\pageno                
                       \rightheadline         
                 \else                        
                       \leftheadline         
                 \fi                          
           \fi}                               
\footline={\hfil}                             
\def\rightheadline{\eightsl\botmark\hfill\eightbf\folio}
\def\leftheadline{\eightbf\folio\hfill\eightsl\lhead}


\def\Alph#1{\ifcase#1\or A\or B\or C\or D\or E\or F\or G\or H\fi}
\def\today{\ifcase\month\or January\or February\or March\or
        April\or May\or June\or July\or August\or September\or
        October\or November\or December\fi\space\number\day,
        \number\year}

\input pstricks
\input pst-node
\input pst-plot
\def\fileversion{97 patch 1}
\def\filedate{1997/05/05}
\message{ v\fileversion, \filedate}

\csname PSTfilesLoaded\endcsname
\let\PSTfilesLoaded 

\ifx\PSTricksLoaded \else
  
  \expandafter\input pstricks.tex
\fi

\edef\TheAtCode{\the\catcode`\@}
\catcode`\@=11


\def\TeXtoEPS{%
\ifx\documentclass\@undefined\else\@begindvi\fi
\begingroup\pst@makebox{}\bgroup\ignorespaces}
\def\endTeXtoEPS{%
    \egroup
    \begingroup
      \global\setbox\pst@boxg\box\voidb@x
      \output{\global\setbox\pst@boxg\box\@cclv}%
      \par\hbox{}\penalty-10000
    \endgroup
    \dp\pst@boxg\z@
    \ht\pst@boxg\z@
    \wd\pst@boxg\z@
  \pst@dimg=3pt
  \shipout\hbox{%
    \box\pst@boxg
    \pst@Verb{gsave CM \tx@STV CP newpath moveto
      \pst@number\pst@dimg neg 0 rmoveto clip setmatrix}%
    \vrule width \wd\pst@hbox height \ht\pst@hbox depth \dp\pst@hbox
    \pstVerb{currentpoint grestore moveto}%
    \kern -\wd\pst@hbox
    \unhbox\pst@hbox}%
  \endgroup
  \typeout{PSTricks: Converting TeX box to EPS.}%
  \typeout{\@spaces\@spaces\space\space With dvips, use -E option.}}


\newwrite\pst@epsout
\newwrite\pst@tempout

\def\pst@getbp#1#2#3{%
  \csname psset#1length\endcsname\pst@dimg{#2}%
  \advance\pst@dimg .49bp
  \pst@cntg=\pst@dimg
  \multiply\pst@cntg 5
  \divide\pst@cntg 328909
  \edef#3{\the\pst@cntg}}

\def\psset@bbllx#1{\pst@getbp{x}{#1}\psk@bbllx}
\psset@bbllx{-1pt}

\def\psset@bblly#1{\pst@getbp{y}{#1}\psk@bblly}
\psset@bblly{-1pt}

\def\psset@bburx#1{\pst@getbp{x}{#1}\psk@bburx}
\psset@bburx{1pt}

\def\psset@bbury#1{\pst@getbp{y}{#1}\psk@bbury}
\psset@bbury{1pt}

\def\pst@getboolean#1#2{%
  \def\pst@tempg##1##2\@nil{%
    \ifx t##1\relax\def#2{1\relax}\else\def#2{\z@}\fi}
  \pst@expandafter\pst@tempg{#1}\relax\@nil}

\def\psset@headers#1{%
  \def\pst@tempg##1##2\@nil{%
    \ifx u##1\relax
      \def\psk@headers{1\relax}%
    \else
      \ifx a#1\relax
        \def\psk@headers{2\relax}%
      \else
        \def\psk@headers{\z@}%
      \fi
    \fi}%
  \pst@expandafter\pst@tempg{#1}\relax\@nil}
\psset@headers{none}

\def\psset@checkfile#1{\pst@getboolean{#1}\psk@checkfile}
\psset@checkfile{true}

\def\psset@makeeps#1{%
  \def\pst@tempg{#1}%
  \ifx\pst@tempg\@none
    \def\psk@makeeps{\z@}%
  \else
    \def\pst@temph{all}%
    \ifx\pst@tempg\pst@temph
      \def\psk@makeeps{3\relax}%
    \else
      \def\pst@temph{all*}%
      \ifx\pst@tempg\pst@temph
        \def\psk@makeeps{2\relax}%
      \else
        \def\psk@makeeps{1\relax}%
      \fi
    \fi
  \fi}
\psset@makeeps{new}

\def\psset@headerfile#1{\def\psk@headerfile{#1}}
\psset@headerfile{}

\def\pst@checkfile#1{%
  \ifcase\psk@makeeps
    \@pstfalse
  \or
    \immediate\openin1=#1
    \ifeof1\relax\@psttrue\else\@pstfalse\fi
    \immediate\closein1
  \or
    \immediate\openin1=#1
    \ifeof1
      \@psttrue
    \else
      \typeout{^^J%
      PSTricks warning:^^J%
      !! File `#1' exists and will be erased if in current directory!^^J%
      !! Make `#1'? (y=yes; n=no)}
      \read16 to \pst@tempg
      \if y\pst@tempg\relax\@psttrue\else\@pstfalse\fi
    \fi
    \immediate\closein1=#1
  \or
    \@psttrue
  \fi}

\def\PSTtoEPS{\def\pst@par{}\pst@object{PSTtoEPS}}
\begingroup
\catcode`\%=12
\catcode`\"=14
\long\gdef\PSTtoEPS@i#1#2{"
  \begingroup
    \use@par
    \newlinechar`\^^J"
    \pst@checkfile{#1}"
    \if@pst
      \immediate\openout\pst@epsout=#1
      \def\write@eps##1{\immediate\write\pst@epsout{##1}}"
      \def\addto@pscode##1{"
        \begingroup
          \newlinechar`\ "
          \write@eps{##1}"
        \endgroup}
      \write@eps{"
          \psk@bbllx\space\psk@bblly\space\psk@bburx\space\psk@bbury^^J"
      \ifcase\psk@headers
        \def\pst@tempg{}"
      \or
        \let\pst@tempg\psk@headerfile
      \or
        \edef\pst@tempg{\pst@theheaders,\psk@headerfile}"
      \fi
      \ifx\pst@tempg\@empty\else
        \expandafter\pst@writeheaders\pst@tempg,\@nil
      \fi
      \write@eps{
      \addto@pscode\pst@dict
      \write@eps{
      \ifdim\pstunit=1bp\else
        \pst@dimg=\pstunit\relax
        \write@eps{\number\pst@dimg\space 65781.8 div dup scale}"
      \fi
      \addto@pscode{"
        \tx@STP
        0 setlinecap
        0 setlinejoin
        10 setmiterlimit
        [] 0 setdash
        newpath}"
      \setbox\pst@hbox=\hbox{"
        \def\init@pscode{"
          \write@eps{"
            gsave
            \psk@origin
            \psk@swapaxes
            \pst@number\pslinewidth SLW
            \pst@usecolor\pslinecolor}}"
        \def\use@pscode{\write@eps{grestore}}"
        \let\pst@rawfile\pst@filetoeps
        \def\psclip#1{\pst@misplaced\psclip}"
        \let\endpsclip\relax
        \def\pstextpath@@@[##1](##2,##3)##4{\pst@misplaced\pstextpath}"
        \def\nc@object##1##2##3##4{\pst@misplaced{node connection}}"
        \def\PSTtoEPS@i##1##2{\pst@misplaced\PSTtoEPS}"
        #2}"
      \write@eps{"
        end^^J"
      \immediate\closeout\pst@epsout
    \fi
  \endgroup
  \ignorespaces}

\gdef\pst@writeheaders#1,{"
  \def\pst@tempg{#1}"
  \ifx\pst@tempg\@empty\else
    \write@eps{
    \pst@filetoeps{#1}"
    \write@eps{
  \fi
  \@ifnextchar\@nil{\let\pst@tempg}{\pst@writeheaders}}
\endgroup

\def\pst@filetoeps#1{%
  \begingroup
    \def\do##1{\catcode`##1=12\relax}%
    \dospecials
    \def\addto@pscode##1{\write@eps{##1}}%
    \pst@@rawfile{#1}%
  \endgroup}

\catcode`\@=\TheAtCode\relax


\input epsf
\input amssym.tex

\def\cosTD{4.1}
\def\cosTd{\cosTD}
\def\perPar{5.1}
\def\primCell{3.1}
\def\novFractalDisc{II}
\def\novFractal{III}
\def\cosSk{IV}
\def\cosZ{V}
\def\cosN{X}
\def\parSk{XI}
\def\parZ{XII}
\def\parN{XVIII}
\def\Boundaries{\cosSk}
\def\zoneTFF{4.4}
\def\zoneTFFO{I.1}
\def\zoneTFFN{I.9}
\def\crPtExchange{4.3}
\def\boxCounting{2.3}

\voffset .5cm
\hoffset -1cm
\null
\vskip 1.truecm

\title{Numerical analysis of the Novikov problem}
\vskip -.5truecm
\title{of a normal metal in a strong magnetic field}
\vskip -.5truecm
\author{Roberto De Leo}{rdl@math.umd.edu}{University of Maryland}
{College Park, MD 20742, USA}

\vskip 1cm

$$\vbox{\hsize 15.cm\noindent
{\bf Abstract}. {\sl We present the results of our numerical exploration of the 
fractal structure found by S.P. Novikov in the problem of the behaviour of 
magnetoresistance in a normal metal under a strong magnetic field. 
The case we discuss in this paper is the simplest non-trivial one, namely the case 
of 2 Fermi Surfaces that cut the brillouin zone along of the coordinate axes
(i.e. Fermi surfaces have genus 3).}
}$$

\section\void{Introduction}

Remarkable topological properties of the problem of asymptotics of orbits of 
quasi-momenta in the dual lattice of a normal metal under a strong magnetic field 
have been noticed by S.P. Novikov in early eighties [Nov82].\par
After analysis of the system behavior for magnetic fields close to rational [Zor84]
and in ``generic position'' [Dyn93a,Dyn97], the following picture has been extracted
by S.P. Novikov (see [NM98] for a review and more bibliography): once a Fermi function
(or a set of {\sl dispersion relations}) has been fixed, on the space of directions of 
the magnetic field is defined a fractal consisting of smooth polygons that generically
have a finite number of points in common.
Every such polygon is labeled by an integer plane (i.e. a Miller index) and to every 
point of it are associated 2 energies. \par
The meaning of these data is the following:
suppose a metal has Fermi Function $\varepsilon$ and Fermi Energy $E_F$ and we want to
know the asymptotic behavior of trajectories of quasi-momenta for some magnetic
field $H$, and be $l = (i,j,k)$, $e_1$ and $e_2$ the Miller index 
and energies associated to $H$. Then the answer is that if $e_1< E_F< e_2$ 
there are open orbits and they are finite deformation of the straight line
of direction $H\times l$, while if $E_F< e_1$ or $e_2< E_F$
all orbits are closed.\par
From these facts it is clear that knowing the zones, their labels and the functions 
$e_1$ and $e_2$ gives us a
complete knowledge of the asymptotic behavior of trajectories. Even in the most 
elementary cases though it is impossible to get analytical expression for 
functions $e_i$ and $l$, so a numerical analysis of the problem is necessary.\par
When $E_F=e_1=e_2$ we can instead get much more complicated ``ergodic-like'' behavior 
of trajectories. For those directions no label is defined in general, unless they
belong to the boundary of some stability zone. 
It is known [DL99] that generically, in the fixed energy picture,
the measure of such directions is zero and that the set has a ``Cantor-like'' fractal 
structure, but nothing is known about their measure in the global picture.\par
The following conjecture has been formulated by Novikov: for a generic set of
dispersion relations, the measure of ``ergodic'' directions in the global picture 
is zero and their fractal dimension is between 1 and 2.
\par
Our numerical study aims to find the first numerical evidence of the existence of
this fractal structure and to evaluate its Minkowski fractal dimension [Fal97] in
the simplest smooth non trivial case, namely the function 
$f(x) = \cos(x)+\cos(y)+\cos(z)$.\par
We also repeat calculations for a piecewise polynomial function with the same 
symmetries, which allows us to work more from the analytical point of view.\par

\section\void{The idea behind the algorithm}

Let us start explaining what we want our algorithm to do.\par
Our ingredients are a function $f$, smooth or piecewise smooth, a non critical value 
$c$ of the function that give rise to the smooth surface $M^2_c=f^{-1}(c)\in \tT$ and 
a direction (magnetic field) $H\in \bR\hbox{P}^2$ (we disregard here any effect 
concerning magnetic breakdown; we just assume that our magnetic field is strong 
enough to give rise to the phenomenon, i.e. it is at least of the order of 
$\sim10^4$ Gauss, and not strong enough to deform the Fermi Surface, so that the only 
free parameter left is its direction).\par
The goal is to get the Miller index associated with $H$, i.e. the homology class 
of the 2-tori (if they exist) on which lie the open orbits corresponding to $H$
(see [NM98] for details). 
In other words we must find somehow three integer numbers 
$l=(l,m,n)\in H_2(\tT,\bZ)$ 
that represent the integer homology class of some 2-torus embedded in $\tT$.\par
To be able to understand the way to get these numbers, we make the following
consideration: assuming we fix some rational magnetic field $H$ such that
$e_2(H)\neq e_1(H)$, our surface $M^2_g$ will be split 
into an even number of 2-tori connected through cylinders of closed orbits, a 
configuration which is homeomorphic to the one shown below.
In general a surface of genus $g$ will give rise to at most $g-1$ cylinders that
in turn will separate at most $g-1$ or $g-2$ 2-tori when $g$ is respectively
odd or even.\par
It is clear that if we cut the cylinders with any pair of 2-tori
with the same homology class of the ones on which the open orbits lie we get a 
bordism between the sets of cycles cut by them on cylinders. 
In other words, such a section would give us a set of 1-cycles, non trivial in
$M^2_c$, so that the sum $z$ of their homology classes does not depend on the height 
at which we choose to intersect.\par
The key point for our algorithm is that there is a 1-1 correspondence
between this homology class $z\in H_1(M^2_c,\bZ)$ and the homology class
$h\in H_2(\tT,\bZ)$ of the 2-tori.\par
Let us call $i:M^2_c\to\tT$ the embedding of the surface in the 3-torus and
$i_*:H_1(M^2_c,\bZ)\to H_1(\tT,\bZ)$ the induced homomorphism on the first 
homology groups. It is straightforward to observe that $z\in\ker(i_*)$ and
that the cycles that lie on the 2-tori, i.e. the
ones that are sent by $i_*$ in the 2-dimensional sublattice of $H_1(\tT,\bZ)$
corresponding to $h$, have intersection number 0 with $z$ and that, on
the contrary, all cycles that have an intersection number different from 0 
with $z$ cannot lie on a single 2-torus.\par
It follows hence that the value of $h$ can be found getting the images
by $i_*$ of all cycles that have number of intersection 0 with $z$, or equivalently
that are symplectically orthogonal to $z$ with respect to the natural structure of
symplectic vector space of $H_1(M^2_c,\bR)$.\par
For example, let us see what happens in the cases we investigated
numerically so far: both functions we studied, in the range of
values $(-1,1)$, give rise to a genus-3 surface embedded in $\tT$ with rank 3 and
their three handles cut symmetrically (see figures \cosTD\ and \perPar) the six 
sides of the cube $[0,1]^3$ (that we use as a model for $\tT$). In particular we are 
going to have just two cylinders that separate two 2-tori, so every section by a 
2-torus parallel (i.e. with the same homology class) to them will cut exactly one 
cylinder and give rise to exactly one homology class $z\in\ker i_*$.\par
If we choose as basic cycles the ones coming naturally from the embedding, as shown 
in picture \perPar, and call them and their canonical duals respectively 
$a_x$, $a_y$ and $a_z$ and $b_x$, $b_y$ and $b_z$, we see that $i_*$
sends the $a_i$'s respectively to $(1,0,0)$ , $(0,1,0)$ and $(0,0,1)$ and 
sends the $b_i$'s to $(0,0,0)$. The cycle $z$ is then a linear combination of 
$b_i$'s with integer coefficients, $z=l b_x + m b_y + n b_z$, and the
cycles sent in $h$ by $i_*$ are the ones such that $<c,z>=0$, 
$c = l^\prime a_x+ m^\prime a_y + n^\prime a_z$.\par
$$
\vbox{\halign{\hfill#\hfill\cr
\pspicture(0,0)(8,8)

\psline(2,7.5)(0,7.5)(-2,5.5)(8.5,5.5)(10,7.5)(7.57,7.5)
\psline[linestyle=dashed,dash=3pt 2pt](2,7.5)(3,7.5)
\psline[linestyle=dashed,dash=3pt 2pt](7.57,7.5)(6.43,7.5)
\psline(6.43,7.5)(3,7.5)
\psline(.05,4)(-1,4)(-3,2)(8.5,2)(10,4)(6.1,4)
\psline[linestyle=dashed,dash=3pt 2pt](.05,4)(.95,4)
\psline(.95,4)(3.57,4)
\psline[linestyle=dashed,dash=3pt 2pt](3.57,4)(4.43,4)
\psline(4.43,4)(4.9,4)
\psline[linestyle=dashed,dash=3pt 2pt](4.9,4)(6.1,4)

\psellipse(.5,6.2)(.45,.2)
\parametricplot[linestyle=dashed,dash=3pt 2pt]{0}{180}{t cos .45 mul .5 add t sin .2 mul 2.8 add}
\parametricplot{180}{360}{t cos .45 mul .5 add t sin .2 mul 2.8 add}
\psline[linestyle=dashed,dash=3pt 2pt](.05,6.2)(.05,5.5)\psline(.05,5.5)(.05,2.8)
\psline[linestyle=dashed,dash=3pt 2pt](.95,6.2)(.95,5.5)\psline(.95,5.5)(.95,2.8)

\parametricplot[linestyle=dashed,dash=3pt 2pt]{0}{180}{t cos .5 mul 2.5 add t sin .16 mul 7 add}
\parametricplot{180}{360}{t cos .5 mul 2.5 add t sin .16 mul 7 add}
\psline(2,7)(2,8)\psline[linestyle=dashed,dash=3pt 2pt](2,3.5)(2,2)\psline(2,2)(2,1)
\psellipse(2.5,3.5)(.5,.16)
\psline(3,7)(3,8)\psline[linestyle=dashed,dash=3pt 2pt](3,3.5)(3,2)\psline(3,2)(3,1)

\psellipse(4,6)(.43,.15)
\parametricplot[linestyle=dashed,dash=3pt 2pt]{0}{180}{t cos .43 mul 4 add t sin .15 mul 2.5 add}
\parametricplot{180}{360}{t cos .43 mul 4 add t sin .15 mul 2.5 add}
\psline[linestyle=dashed,dash=3pt 2pt](3.57,6)(3.57,5.5)\psline(3.57,2.5)(3.57,5.5)
\psline[linestyle=dashed,dash=3pt 2pt](4.43,6)(4.43,5.5)\psline(4.43,5.5)(4.43,2.5)

\psellipse(5.5,7)(.6,.2)
\parametricplot[linestyle=dashed,dash=3pt 2pt]{0}{180}{t cos .6 mul 5.5 add t sin .2 mul 3.5 add}
\parametricplot{180}{360}{t cos .6 mul 5.5 add t sin .2 mul 3.5 add}
\psline[linestyle=dashed,dash=3pt 2pt](4.9,7)(4.9,5.5)\psline(4.9,5.5)(4.9,3.5)
\psline[linestyle=dashed,dash=3pt 2pt](6.1,7)(6.1,5.5)\psline(6.1,5.5)(6.1,3.5)

\parametricplot[linestyle=dashed,dash=3pt 2pt]{0}{180}{t cos .57 mul 7 add t sin .2 mul 6.5 add}
\parametricplot{180}{360}{t cos .57 mul 7 add t sin .2 mul 6.5 add}
\psline(6.43,6.5)(6.43,8)\psline(7.57,6.5)(7.57,8)
\psellipse(7,3)(.57,.2)
\psline[linestyle=dashed,dash=3pt 2pt](6.43,3)(6.43,2)\psline(6.43,2)(6.43,1)
\psline[linestyle=dashed,dash=3pt 2pt](7.57,3)(7.57,2)\psline(7.57,2)(7.57,1)

\endpspicture\cr
\caption\cylinders{an example of the splitting in 2-tori and cylinders 
of a surface of genus 7 under the action of a constant magnetic field.}\cr
}}
$$

It is then clear now that the homology class $h$ is represented in $H_2(\tT,\bZ)$
by the same triple of integers that represents $z$ in $H_1(M^2_c,\bZ)$,
so in this particular case it is enough to find out the intersection number
of $z$ with the $a_i$'s to get $h$.\par

\section\void{The NTC library}

After the previous discussion it is clear which capabilities we expect from
the software we are going to use to perform the numerical analysis.
It must be able to deal with the topology of a surface, i.e. it must
have the possibility of dealing with simplexes of dimension 0, 1, 2 and 3,
and it must be able to perform topological operations like getting a simplicial
decomposition of the level set of a function of three variables (to get the Fermi surface 
$M^2_c$), intersecting two simplicial complexes (to get the 1-dimensional leaves 
on $M^2_c$), identifying closed curves in $\tT$ (i.e. it must be able to deal with the 
periodic boundary conditions that identify $[0,1]^3$ with $\tT$), 
evaluating intersection numbers 
between 2 cycles on a surface and finding the homology class of loops in $\tT$.\par
When we started working at this project, after thorough search on 
the InterNet we found several C++ libraries able to deal with the topology
of 3- and lower-dimensional objects through simplicial decompositions
(``meshes'' in computer jargon). None of them of course directly implements 
the specific functions we needed, so we decided to write a C++ library on top
of one of the preexisting ones to implement the complex 
topological functions we needed and tried to make the code as much reusable 
as possible as it seems that such a library could be useful in the future also 
for different numerical topological problems. We called our library Novikov 
Torus Conjecture library (NTC).\par
After an accurate examination of all libraries available we chose to use
the library VTK (Visualization ToolKit, http://www.kitware.com). 
The main reasons for our choice, aside from the fact that it is free, are 
the availability of its source code, the fast rate at which is developed 
and the existence of a very active mailing list about VTK-related problems 
and solutions. Moreover, as VTK was intended primarily as a visualization
tool based on the standard C library OpenGL by SGI (http://www.opengl.org), 
it easily allows us to visualize our surfaces and cycles, making much
easier the debugging process.\par
Let us now explain in detail how the algorithm works.\par 
First of all it is important to point out that, as every computer can basically deal 
just with integer numbers, we are able just to explore what happens for
1-rational magnetic fields. This is not a big restriction as the homology
class we look for is locally constant with respect to the direction of the magnetic 
field and so, close to every $H$, there is a rational one that induces
the same homology class.\par
Let us then fix some 1-rational $H$. 
Its critical points generically will be of ``figure eight'' type, i.e.
its tails close up to a pair of loops. We call these critical points ``fully open'', 
``half-closed''
and ``fully-closed'' using a terminology coming from $\tR$, or equivalently 
depending on whether in $\tT$ both loops are  non trivial, one is homotopic to 0 
or they are both trivial. \par
The first thing to do is to check whether 
it gives rise to the generic situation or not, i.e. whether open orbits lie on 
2-tori or not. 
For simplicity we will just analyze the 
case of genus 3 surfaces, in which just two 2-tori and two cylinders can appear,
that are the only cases studied numerically by now.\par
As we said, the 2-tori components are separated by cylinders
of closed orbits that have as basis a pair of ``half-closed'' critical saddle 
points (see for example figure \primCell), i.e. saddles in which two of the 
four tails are connected so that the loop they form is homotopic to 0 in $\tT$. 
Our strategy then is 
to find numerically the set of all critical points, single out the saddles
that play role in the topology of the surface (we disregard all saddles that
are associated to a center; they are easy to spot as they are half or fully closed
and form a loop homotopic to 0 in the surface) and count how many of them are half 
closed.\par

$$
\vbox{\halign{\hfill#\hfill\cr
\pspicture(0,0)(10,6.5)
\rput(5,3.5){\epsfxsize=13.cm\epsfbox{PrimitiveCell.epsf}}
\endpspicture\cr
\caption\primitiveCell{an example of the level set of the Fermi function
restricted to a basic cell of a 1-rational foliating plane.
To the critical saddle point are attached
two loops, just one of them homotopic to 0 (in $\dT\subset\tT$).}\cr
}}
$$

If at least three of them, and so all four, are half closed, then we evaluate 
the homology 
class in the surface of any of the loops and this would give us exactly the 
homology class of the 2-tori of open orbits. \par
If instead one, and so at least two, of them is fully closed or fully 
open, then there cannot be any rank-2 2-torus; there could be still open orbits
but they will fill rank-1 tori, i.e. cylinders, and these open orbits will
disappear for any generic small enough perturbation, so we do not register 
any homology class for these directions.\par
This algorithm is implemented through the construction of several classes.
The main classes are ntcFoliation and its subclass ntcPlaneFoliation, that 
contain all parameters (Fermi function, energy, magnetic field and resolution 
of samplings) and functions able to produce and classify the critical points of 
the foliation and to get the critical leaf
(our terminology comes from the universal covering $\tR$: in $\tT$ for 1-rational 
magnetic fields critical leaves will be always fully closed,  but the ones leading 
to open orbits will have just one of the loops homotopic to 0 in $\tT$ so in the 
covering such saddle is half-open).\par
For several reasons we implemented two different ways to get the critical leaves.
One works by obtaining the level lines of the Fermi function restricted to the 
plane perpendicular to $H$ passing through the singular point. To get
the full picture of the intersection we restrict our sampling to an opportunely
chosen parallelogram spanned by a $\bZ^2$ basis of the 2-dimensional lattice
given by the intersection of $\bZ^3$ with the plane perpendicular to $H$,
so that we get a picture that glues nicely on the boundary.\par
This procedure tends to need too much RAM when the components of $H$ get 
big, say around 400, because the area of the basic parallelogram tends to get
too big. We implemented therefore a second way that needs just the simplicial
decomposition of the Fermi surface in the cube $[0,1]^3$. 
The critical leaf
now is obtained starting cutting by the plane passing through the critical 
point and then following the loop. When the trajectory reaches the boundary
of the cube the coordinates of the equivalent point are evaluated and a new
plane is taken passing through that point. The process stops when the loop
comes back for the second time to the critical point (as there are two critical
loops for every critical point).\par

$$
\vbox{\halign{\hfill#\hfill\cr
\pspicture(0,0)(6,6.2)
\rput(3,3.5){\epsfxsize=6.cm\epsfbox{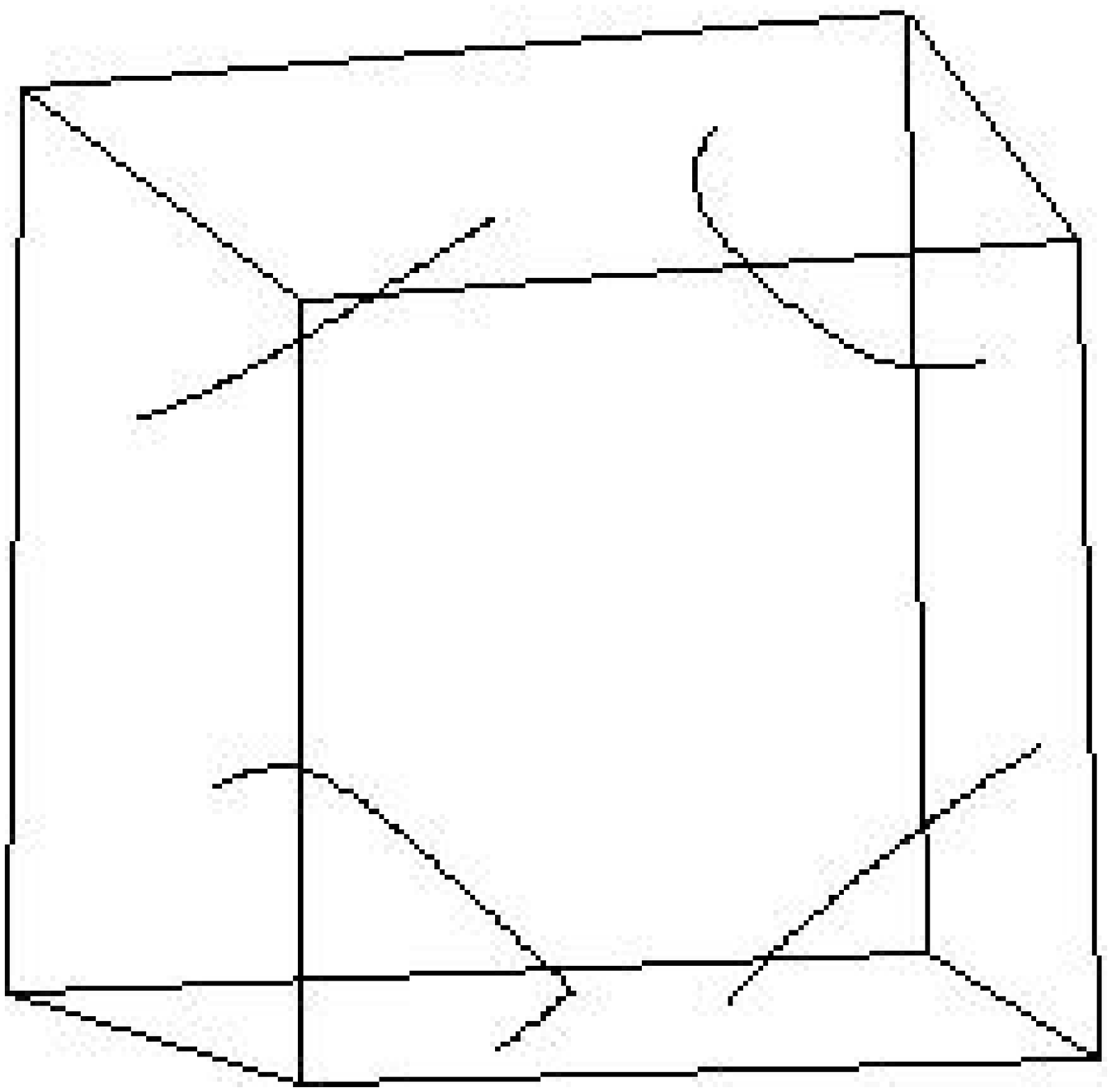}}
\endpspicture\cr
\caption\primitiveCell{an example of the level set of the Fermi function
obtained through the second kind of algorithm implemented. Just the 
critical loop homotopic to 0 (in $\tT$) is shown.}\cr
}}
$$

In both cases the critical leaf is at the end copied in an object
of the class ntcPrimitiveCell that contains methods to deal with loops in
$\tT$. If just one of the two loops is homotopic to zero in $\tT$, then 
the methods for evaluating the intersection numbers of that cycle with
the 3 cycles in the kernel of $i_*$ are called (they are contained in the
class ntcImplicitFunction that contains all data about the function)
and the result gives automatically the searched homology class.\par
A few other complementary class are also implemented to deal with leaves
and functions. The complete documentation for the NTC library together with 
the source code is available at the InterNet address 
http://www.math.umd.edu/$\sim$rdl/ntc/.

\section\cosSec{Study of the trigonometric function}

The function $f(x,y,z) = \cos(2\pi x)+\cos(2\pi y)+\cos(2\pi z)$ is the 
simplest trigonometric function that gives rise to a non trivial
(i.e. rank 3) embedding of a surface in $\tT$ and the only one that had
been studied so far.\par
The only critical values of $f$ are $\pm3$ and $\pm1$, so all level sets 
$M_{_E}=f^{-1}(E)$ are homeomorphic to spheres for $E\in(-3,-1)\cup(1,3)$. 
The level set $M_0$ shown in figure \cosTd\  shows that for $E\in(-1,1)$ 
all level sets are genus-3 surfaces embedded with rank 3 in $\tT$.\par
In particular this means that every generic foliation of $M_E$ will
have at least four saddle points, and that all saddles but four will be
associated to some center and hence will be homotopic to 0 in the surface.
These critical points, that we call ``topological'' as their origin is due
to the topology of the surface and not to the particular embedding, are
at the base of two cylinders that divide two 2-tori for every generic 
direction of the magnetic field.\par
Each level surface of this function is invariant under the symmetry 
group of the cube; this action in turn induces an action on $\dS$ under which
the fractal picture is invariant, so it is enough for us to analyze its
structure in one of the 48 domains in which the action subdivides $\dS$.\par
In the projective chart of $\dS$ corresponding to the plane $z=1$, one of 
these domains is the triangle $x\leq y\subset[0,1]^2$, so we will refer just 
to the square $[0,1]^2$ as our ``phase space'' from now on. 
On this square the picture of stability zones will be symmetric with respect to 
the diagonal, fact that will be used as consistency check of our algorithm.\par

$$
\vbox{\halign{\hfill#\hfill\cr
\pspicture(0,0)(8.6,8)
\rput[lb](0,0){\epsfbox{cos3D.epsf}}
\endpspicture\cr
\caption\cosTd{the surface $M_0$ cut by a plane passing through a critical point.}\cr
}}
$$

Another symmetry, due to the fact that the cosine is an even function, 
implies that all level sets are symmetric with respect to the origin.
It follows that the four topological saddles are divided in two 
symmetrical pairs that define one cylinder each. In fact we can assign
to every closed (in $\tR$) orbit a plus or minus sign, according to the fact 
that it bounds a region where $f$ assumes respectively values smaller or bigger 
than the one assumed on the loop (these two different kinds of loop are 
called ``electrons''
and ``holes'' in physics literature). \par
This sign is invariant by homotopy,
so the same sign is associated to the whole cylinder and is shared by
the two critical loops at the two bases. As the symmetry respect to the
center do not switch this sign, it is clear that every pair of symmetric
critical points defines one of the two cylinders.\par 
Finally, the identity $\cos(2\pi x ) = - \cos\left[2\pi (1/2-x) \right]$ induces 
a symmetry between different level surfaces, namely 
the surface $M_c$ is obtained from $M_{-c}$ through a translation and a
reflection respect to the origin. As the foliation $p_iH^i=const$ is invariant 
by these two operations, it is clear that the existence of open orbits
at energy $c$ implies the existence of open orbits at energy $-c$, so that 
the interval for which any direction gives rise to open orbits (that is
closed, connected and non empty by [Dyn97]) has the form $[-E,E]$.\par
The surface $M_0$ hence plays a very special role, as at energy $c=0$ every 
direction gives rise to open orbits and so every ``stability zone'' reaches 
here its biggest size.
This means that to study the fractal on $\dS$ corresponding to this function
is enough to study the level $c=0$, while in general it would be needed to 
check several different energies for every direction of $H$ to find
which homology class, if any, is associated to it.\par
Moreover, this means that at every energy different from $0$ there is no 
common point between boundaries of different zones, as every zone gets
strictly smaller at every change of energy. In the limit for the energy
that goes to -1 or 1 all zones tend to disappear as above 1 or below -1 
the level surface of $f$ is a sphere.\par
Let us now examine in more detail the case of 0 energy: it is easy to verify
that this surface has curvature everywhere negative except in the eight
points $(\pm.5,\pm.5,\pm.5)$ in which is 0. This means that for every
direction different from $(\pm1,\pm1,\pm1)$ we will have exactly four
critical points, all of saddle type because of the topological constraints.\par
The analytical expression of the critical points for a generic $E\in(-1,1)$ is
very complicated but it gets much simpler in the most interesting case, namely
$E=0$. Their expression in cartesian coordinates $(a,b)\in[0,1]^2$ is:

$$\vbox{
	\halign{$#$\hfill&$#$\hfill\cr
	x_1(a,b) &= \displaystyle\frac{1}{2\pi}\sin^{-1}\left( a \alpha\left(a,b\right) \right)\cr
	\noalign{\vskip.2cm}
	y_1(a,b) &= \displaystyle\frac{1}{2\pi}\sin^{-1}\left( b \alpha(a,b) \right)\cr
	\noalign{\vskip.2cm}
	z_1(a,b) &= 
	\cases{
		\eqalign{\frac{1}{2}-\frac{1}{2\pi}\sin^{-1}\left( \alpha(a,b) \right)\;&,\,\,a\leq b\cr
	       \phantom{\frac{1}{2}-}\frac{1}{2\pi}\sin^{-1}\left( \alpha(a,b) \right)\,&,\,\,a\geq b\cr}\cr}\cr
	\noalign{\vskip.4cm}
	(x_2,y_2,z_2) &= \displaystyle(\frac{1}{2},\frac{1}{2},\frac{1}{2}) - (x_1,y_1,z_1)\cr
	\noalign{\vskip.4cm}
	(x_3,y_3,z_3) &= \displaystyle(\frac{1}{2},\frac{1}{2},\frac{1}{2}) + (x_1,y_1,z_1)\cr
	\noalign{\vskip.4cm}
	(x_4,y_4,z_4) &= \displaystyle(1,1,1) - (x_1,y_1,z_1)\cr
	\noalign{\vskip.4cm}
	\alpha(a,b)&= \displaystyle \sqrt{\frac{2\sqrt{ a^4+b^4+1-a^2b^2-a^2-b^2 } - ( a^2+b^2+1 ) }{a^4+b^4+1-2a^2b^2-2a^2-2b^2}}\cr
	\noalign{\vskip.2cm}
}}$$

It is straightforward to realize that the averaged Euler characteristic 
$\chi_{H}(c)=<H, \vec\gamma(c)>$,
where $\vec\gamma(c)=\sum w_i x_i(c)$ is the sum over all critical points
weighted by the Dynnikov index $w_i$ equal to the ``hamiltonian'' index of the 
critical point (as 0 of the 1-form, see [Dyn97]) times  $<\nabla_{_{\!x_i}}\!f,H>$, 
is identically 0 for $c=0$.\par
This fact is 
also clear from the relation $\chi_{H}(c)=\sum h_+ - \sum h_-$, i.e.
the averaged Euler characteristic is equal to the sum of the height of
cylinders of ``positive'' closed leaves (the ones on which the gradient 
points to the exterior of the loop) minus the height of cylinders of 
``negative loops''.\par
By the symmetry at $c=0$, that exchanges ``electrons'' with ``holes'',
i.e. positive cylinders with negative ones, it is clear that the sum is
zero, while it is negative for $c>0$ and positive for $c<0$. This 
corresponds to the fact that all ``ergodic'' or ``non generic'' 
directions appear just at energy 0, as the nullity of the averaged
Euler characteristic is a necessary condition for the appearance of these 
directions.\par
That there could be no ``ergodic r\'egime'' for energies different from 0 was 
also clear from the fact that all energy intervals
$[e_1(H),e_2(H)]$ for which open orbits exist are of the
form $[-e,e]$: ``ergodic'' directions correspond to the case of length
zero of this interval, that in this case implies $e_1=e_2=0$.\par
Now let us see what is possible to do ``by hand'' about stability zones
at energy 0. As we have the explicit analytical expression for 
all critical points we can use the following procedure: first of
all we make sure somehow that a direction $(a,b)$ is ``generic'', i.e. it is 
inside some stability zone, for example looking at the plane section generated
by the NTC library or by any computer algebra program like Mathematica
and verifying that just one of the loops is homotopic to 0 (at energy
0 it is enough to examine just one of the critical points because of
the symmetry).\par
Then we choose one critical point, say $p_1=(x_1,y_1,z_1)$, 
inside the cube $[0,1]^3$ and follow ``vertically'' the cylinder of closed 
orbits until we reach the second base point. As we observed before, the 
second base point $p$ must be its symmetrical respect to the origin,
namely the one we called $p_4$, so in the covering its coordinates will 
be of the form $p_4+(l,m,n)$. Equivalently, going from $p_1$ to $p_4$
inside the cylinder and coming back to $p_1$ through the segment that 
joins them inside the cube will produce a loop of homology class $(l,m,n)$
in $\tT$.
As at the boundary of a zone both cylinders have height 0, i.e. the two bases
belong to the same leaf, it follows that the boundary of any zone is 
a subset of the curves 
$\{<H, p_1 - p_4 - (l,m,n)> = 0\}_{(l,m,n)\in\tZ}$.\par
By the topological stability of curves homotopic to $0$, this triple of integers 
depends continuously on the magnetic field, so it is locally constant. The 
number of different triples inside a single stability zone determines the 
number of sides of the zone as shown in figure \zoneTFF.\par
The cylinder identified by $p_1$ and $p_4$ will disappear either when its
height goes to 0 or when it gets substituted by a new one:
in the first case it means that we reached the boundary of the stability
zone.

$$
\vbox{\tabskip=25pt\halign{\hfill#\hfill\cr
\pspicture(0,0)(9,5)
\rput(4.5,3){\vbox{\halign{\hfill#\hfill&\hfill#\hfill&\hfill#\hfill\cr
	\epsfxsize=4.5cm\epsfbox{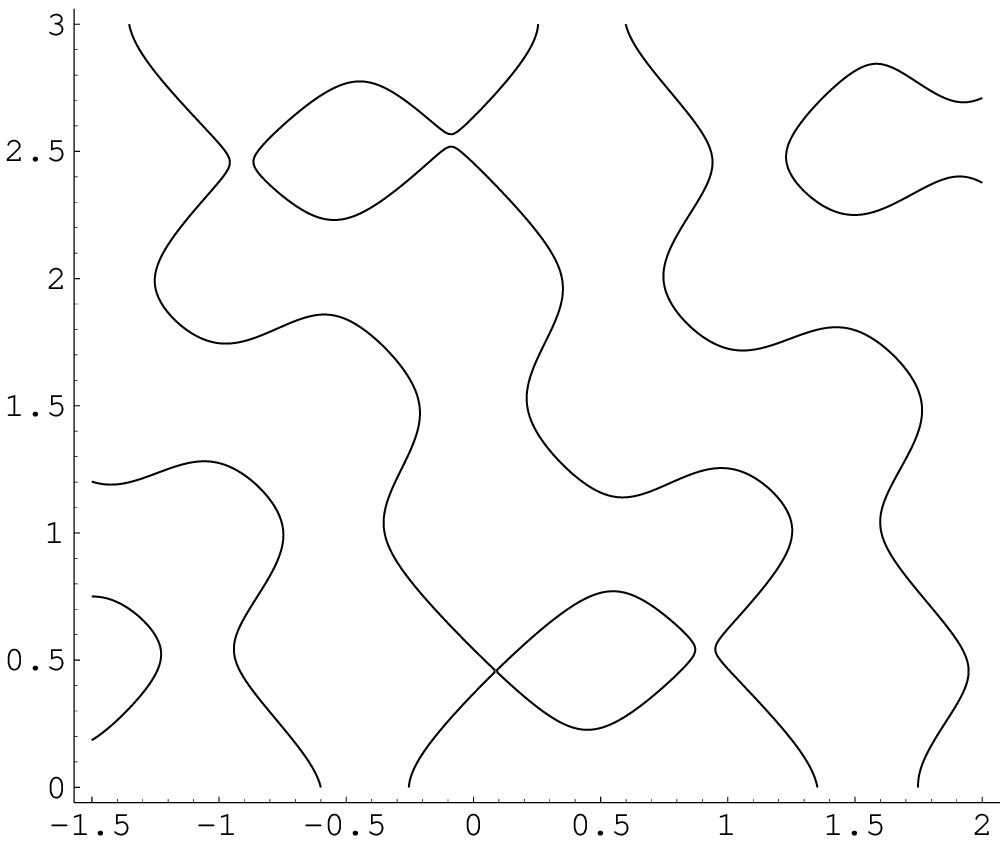}
&
	\epsfxsize=4.5cm\epsfbox{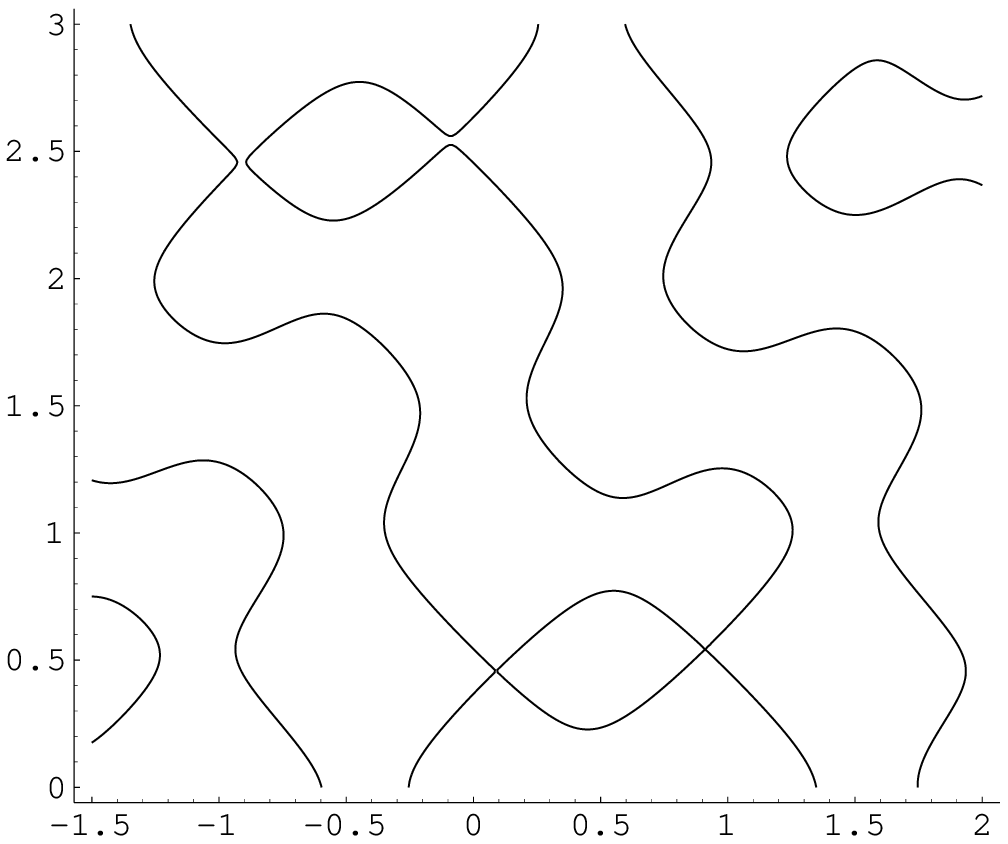}
&
	\epsfxsize=4.5cm\epsfbox{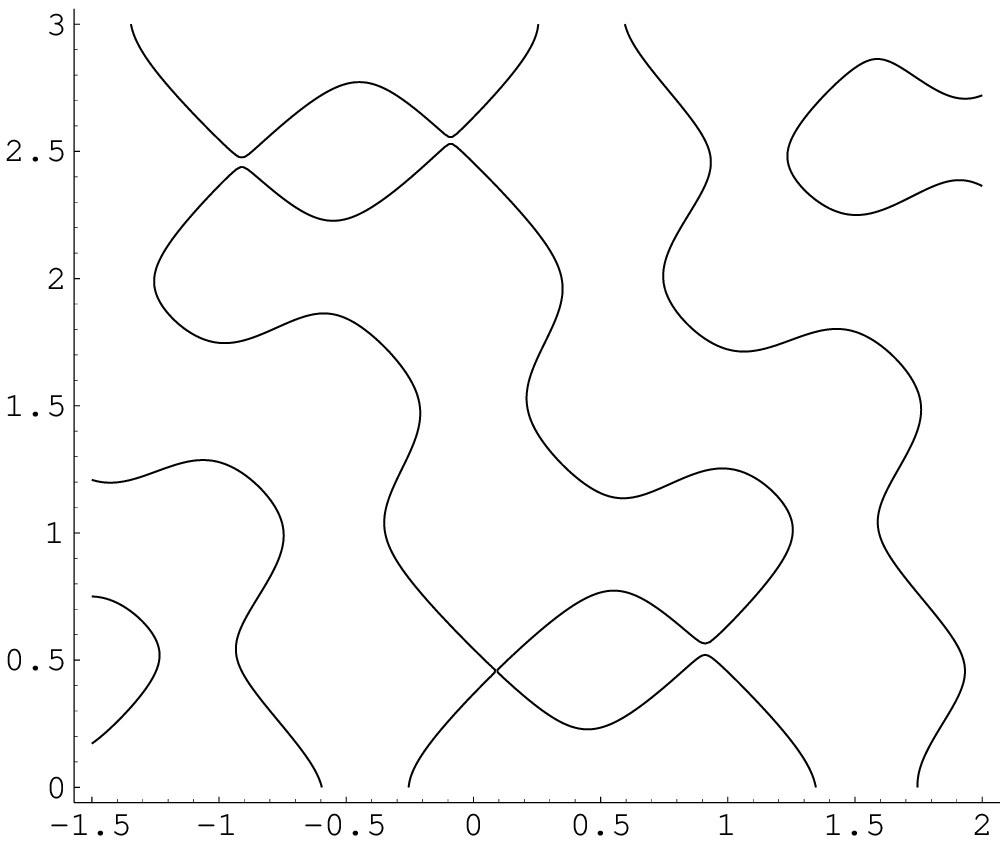}\cr
	\noalign{\vskip .1cm}
	$H = (.53, .268,1)$
&
	$H \simeq (.5352, .268,1)$
&
	$H = (.537, .268,1)$\cr
}}}
\endpspicture\cr
\caption\cilExchange{ the disappearence of a cylinder at the boundary of the 
stability zone (0,0,1): on the left the cylinder has non-zero height, in the 
central picture the two basis collapse one over the other and in the third one 
the cylinder has disappeared, substituted by a new one relative to the stability 
zone (1,2,4).}\cr
}}
$$

In the second case it happens that either two different cylinders collide and 
mutually
exchange one of their bases or a single cylinder collides with itself and the base
point is exchanged with one equivalent to it but in a different position (see figure 
\crPtExchange\ and tables \zoneTFFO-\zoneTFFN). 

$$
\vbox{\tabskip=15pt\halign{\hfill#\hfill\cr
\pspicture(0,0)(9,6)
\rput(4.5,3){\vbox{\halign{\hfill#\hfill&\hfill#\hfill&\hfill#\hfill\cr
	\epsfxsize=4.8cm\epsfbox{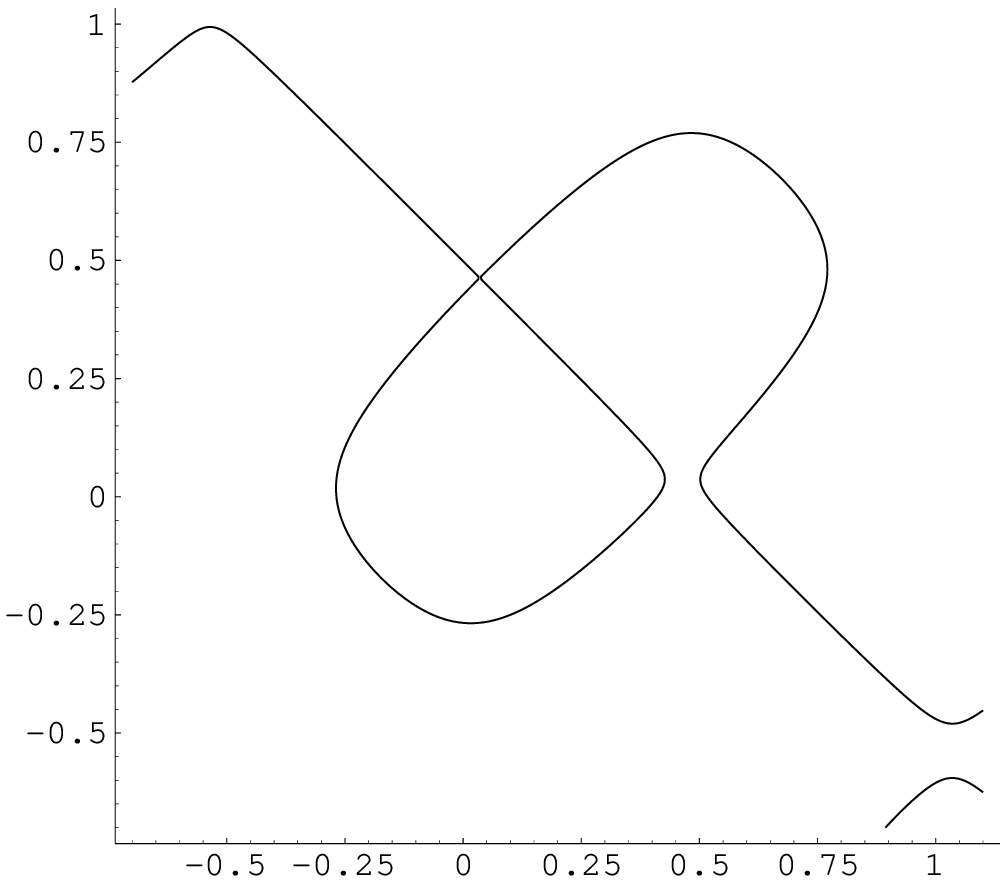}
&
	\epsfxsize=4.8cm\epsfbox{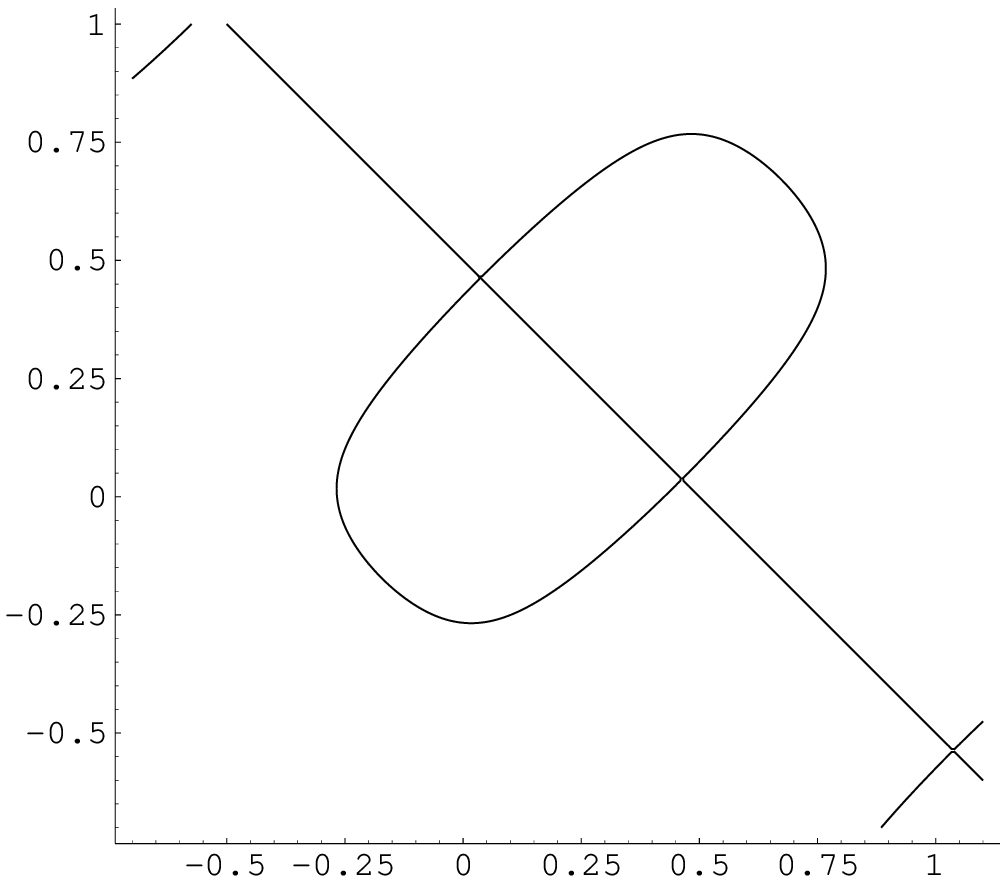}
&
	\epsfxsize=4.8cm\epsfbox{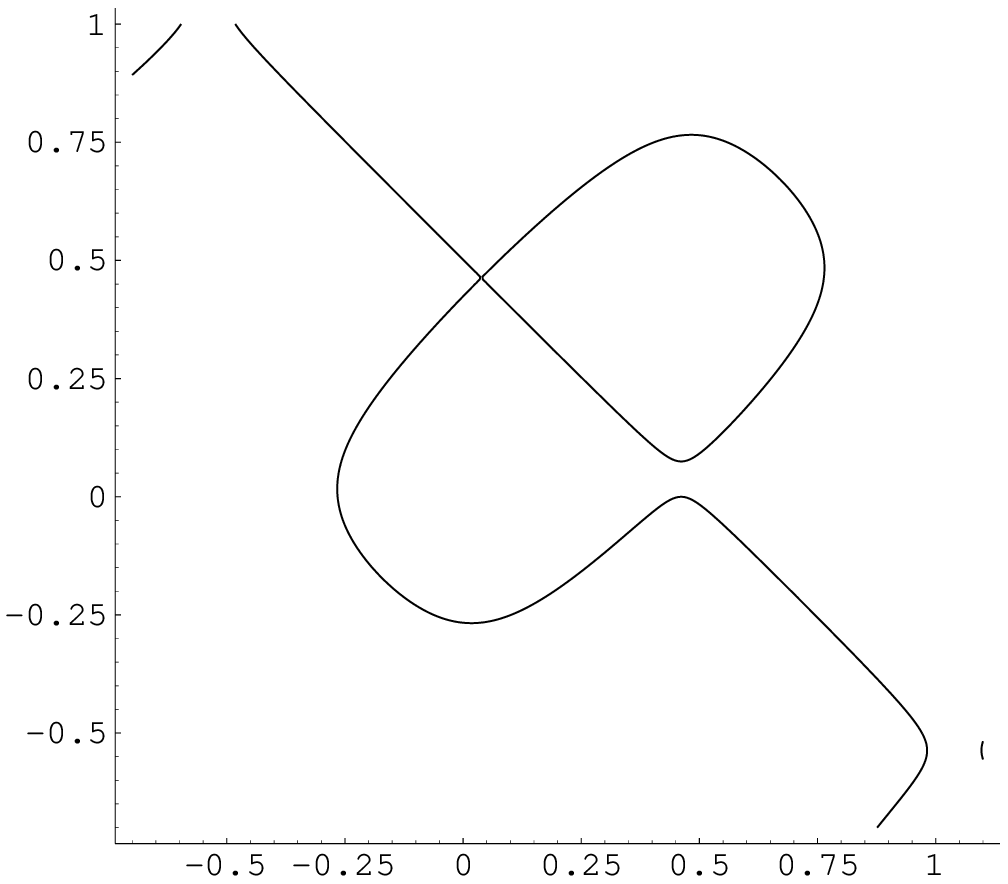}\cr
	\noalign{\vskip .1cm}
	$H=(.22,.23,1)$
&
	$H=(.23,.23,1)$
&
	$H=(.24,.23,1)$\cr
}}}
\endpspicture\cr
\caption\cilExchange{a change of cylinder inside a stability zone. On the left
is shown a critical leaf at the base of a cylinder, the
critical point is $p_1\simeq(0.035,0.463,0.25)$. At the opposite base lies the 
critical point $p=p_4+(0,0,1)$. The middle picture shows what happens at the
boundary between the two stability zones of cylinders, namely 
the point $p_1$ has a saddle connection with $p_2$. The picture on the right
shows the base of the new cylinder. At one base still lies 
the point $p_1$ but at the opposite one now lies $p_4+(1,1,0)$.}\cr
}}
$$

In picture \zoneTFF\ we show what happens in case of the zone $(2,4,5)$:
there are three different kinds of cylinder, labeled by $(-3,3,-2)$,
$(0,0,1)$ and $(-4,2,-1)$, so the zone is a triangle divided inside in
three sub-zones. At the boundary between the first and the second sub-zone
the change is determined by the appearance of a saddle connection 
between $p_1$ and $p_2+(-1,2,-1)$, at the boundary between the first and the
third we have an analogous situation between $p_1$ and $p_2+(-3,3,-1)$
and at the boundary between second and third we have instead the appearance
of saddle connection of $p_1$ with itself, precisely with $p_1+(2,-1,0)$.\par
In table \novFractal\ and also in the other pictures with smaller resolution 
it is possible to recognize in many stability zones the
boundaries between sub-zones in which $p_1$ has a saddle connection with 
itself, as in these points the 2-tori filled by open orbits have rank 1
and so these points are not included in the data and the stability zone
is cut by a segment of straight line. It is easy to check that the same
straight line, whose equation is $la+mb+n=0$ for $p_1$ having a saddle
connection with $p_1+(l,m,n)$, cuts several (possibly infinite) zones.\par

$$
\vbox{\tabskip=15pt\halign{\hfill#\hfill\cr
\pspicture(0,0)(9,8.5)
\rput(4.5,4.5){\epsfxsize=8.5cm\epsfbox{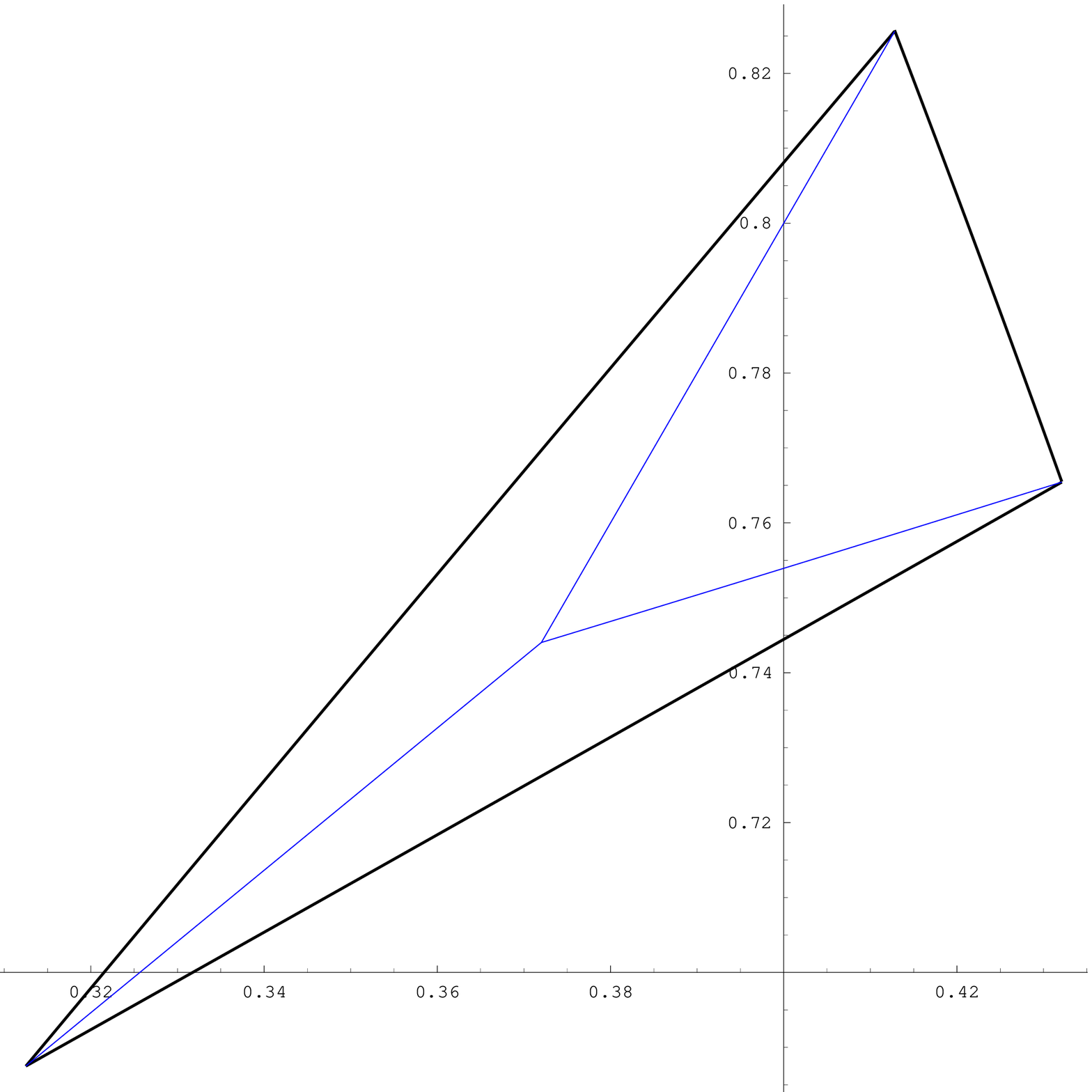}}
\rput(4.6,3.3){\rmU}
\rput(7.1,5.7){\rmD}
\rput(4.2,4.3){\rmT}
\endpspicture\cr
\caption\zoneTFF{the ``cylinder'' structure of the zone $(2,4,5)$. 
Keeping fixed the critical point of one of the bases of the cylinder, say $p_1$,
in subzone \rmU\ in the second base we find the critical point $p_4+(-3,3,-2)$,
in subzone \rmD\ the critical point $p_4+(0,0,-1)$ and in subzone \rmT\ the
critical point $p_4+(-4,2,-1)$. All these three cylinders are formed by closed
loops that have the same homology class in $M_0$, namely $(2,4,5)$ (using
coordinates with respect to the natural base in $\ker i_*$).\par
In tables
\zoneTFFO-\zoneTFFN\ are shown the three kinds of cylinder corresponding to 
the three internal subzones and the way they transform when the magnetic field
direction crosses the internal boundaries.}\cr
}}
$$
\vskip.2cm
These rank-1 2-tori survive longer to energy changes than the rank-2 ones, 
i.e. we still
find them when the rest of the zone has disappeared, but they disappear
for any generic perturbation of $H$.
A further confirmation of the accuracy of our algorithm is given
by the perfect agreement between the segment found analytically for
zone $(2,4,5)$ shown above and the one that is possible to see in table
\novFractal.\par
All techniques described above allow us in principle to find analytically 
all boundaries of stability zones and the boundaries of their sub-zones, 
even though they do nothing to help us finding which 
homology class is associated with them; this quantity of course is anyway easily obtained
through our library. The main problem is that we did not find any way to
put these procedures in any simple algorithm for letting a computer do the job,
so it has to be done ``by hand''.\par
Anyway to be able to get this analytical expression does not seem to be crucial 
in itself: 
with our NTC library we can obtain a good approximation of the interior of any
stability zone by sampling the square $[0,1]^2$ with step $1/N$ in both directions.
In that way we will get for every point $(m/N,n/N)$, $0<m,n\leq N$, the 
homology class of the stability zone it belongs to (if any). It is good though
to have such analytical expressions as they provide a way to double check
the accuracy of our algorithm comparing the interior of the zone found with 
the NTC library with its analytical boundary.\par
We initially run our program with resolution $N=100$ at energies 
$E=0,-.1,-,2,-.3,-.5,-.7,-.9$ and found the pictures we show in tables \cosZ-\cosN.
In table \Boundaries\ are shown the labels associated to the biggest zones 
together with their boundaries found analytically. The boundaries are also
drawn in table \cosZ\ to show the very good agreement with them of numerical data
found using the NTC library. 
After trying several different machines with different operating systems, it 
turned out that the fastest machines available to us were Pentium II Linux 
machines, so we run all our simulations on them. Every sampling with $N=100$ 
resolution takes around 12 hours of CPU.\par
In table \novFractal\ are shown the data found with the NTC library at a resolution
$N=1000$. 
The calculation explored just the upper triangle $b\geq a\subset[0,1]^2$,
it run $\sim3$ weeks on 5 Linux machines with Pentium II CPUs and found 
$\sim3\cdot 10^4$ distinct stability zones. In table \novFractal\ we show just 
the 1066 zones containing at least 10 points and then we extend the picture by 
symmetry to the whole square. In table \novFractalDisc\ we extended the picture 
to the upper half of the sphere by symmetry to show the global pattern of the
fractal.\par 

\subsection{Evaluation of the fractal dimension}

One of the most standard procedures to get the fractal dimension of a set is
to evaluate its ``Box Counting'' dimension [ASY96,Fal97]. To double check our results
we used two different methods to get this estimate.\par 
The first method comes directly from the definition,
namely we divide the square in $2^{2n}$ squares of area $1/2^{2n}$ and count
how many of them we need to cover the fractal (i.e. the white spots in table
\novFractal). Below we show the data for $n=1,\cdots,10$

$$
\vbox{\halign{\hfill#\hfill\cr
\pspicture(0,0)(9,4)
\psaxes[tickstyle=top,Dy=2,dy=.5]{->}(11,4)
\psdots*(1,.4)(2,.86)(3,1.32)(4,1.78)(5,2.21)(6,2.65)(7,3.1)(8,3.51)(9,3.83)(10,3.98)
\rput(10.75,-.4){$n$}
\rput(-1.5,1.75){$\log_2(N_n)$}
\endpspicture\cr
\noalign{\vskip .9cm}
\caption\cylinders{plot of the $\log$ in base 2 of number of squares needed to cover the
fractal with squares of size $2^{2n}$ versus the size scale $n$.}\cr
}}
$$

After we discard the last two terms, that probably we cannot evaluate well enough because
of the finiteness of our resolution,
we find that the slope that minimizes the 
$rms$ in a linear fitting of the above plot is $d\simeq1.78$.\par
The other method we used is the following: after having fixed a number $r>1$ we count for
every $n$ how many zones have area between $r^{-n}$ and $r^{-n-1}$. Let us call this
number $N_n$. Then as $n\to\infty$ the ratio between $\log_r(N_n)$ and $n$ converges,
for fractals for which that dimension is well defined, to the Box Counting dimension 
divided by the dimension of the ambient space [Fal97].\par
The picture below shows the plot in case $r=2$:

$$
\vbox{\halign{\hfill#\hfill\cr
\pspicture(0,0)(9,4)
\psaxes[tickstyle=top,Dy=2,dy=.5,Dx=2,dx=1]{->}(9.75,4)
\psdots*(.5,0)(1,0)(1.5,0)(2,.25)(2.5,.58)(3,.4)(3.5,.65)(4,1.1)(4.5,1.21)(5,1.4)
(5.5,1.64)(6,1.84)(6.5,2.08)(7,2.31)(7.5,2.59)(8,2.94)(8.5,3.62)
\rput(9.45,-.4){$n$}
\rput(-1.5,1.75){$\log_2(N_n)$}
\endpspicture\cr
\noalign{\vskip .9cm}
\caption\cylinders{plot of the $\log$ in base 2 of number of number of zones of area 
between $2^{-n}$ and $2^{-n-1}$ versus the size scale $n$.}\cr
}}
$$

In this case the global behavior is much less linear but is clear that the first points
have no real meaning because there the scale is still too big and it is safe also to
discard the last ones as there we are probably at a scale too small for the resolution 
of our picture. After discarding these boundary points we are left with a plot which
can be well approximated by a linear function with slope $\alpha\simeq.91$.
This suggests that $d\simeq1.82$, which is in very good agreement with the previous 
estimate for the fractal Box Counting dimension of the set of ``ergodic'' directions.

\section\void{Study of the piecewise quadratic function}

Using degree-2 polynomials we can build a function that has the same properties of the 
previous one but is much easier to deal with analytically. 
The function will not be globally smooth as the second derivatives will not glue 
smoothly, but still it will be globally ${\cal C}^1$ and piecewise quadratic.

$$
\vbox{\halign{\hfill#\hfill\cr
\pspicture(0,0)(10,8.5)
\rput(5,4.5){\epsfxsize=8.cm\epsfbox{perPar.epsf}}
\endpspicture\cr
\caption\cosTd{the surface $N_0 = f^{-1}(0)$. The three basic cycles non homotopic to 0 
in $\tT$ are shown.}\cr
}}
$$

In particular it is possible to build a piecewise polynomial function of
degree 2 that allows to evaluate the espression of all critical points at
every energy, so that we will be able in principle to verify the agreement
of our algorithm with every zone at energies different from zero and to find
analytical expressions for topological quantities that depends on them like the
averaged Euler characteristic.\par
The function we used is the following:
$$
f(x,y,z) = F(x) + F(y) + F(z)
\;,\;\;\;
F(x) = \cases{
	\phantom{-}8(2[x]-1)[x]\phantom{([x]-1)}\!\!\!,\;[x]\in[0,.5]\cr 
	-8(2[x]-1)([x]-1)\phantom{[x]}\!\!\!,\;[x]\in[.5,1]\cr
	}
$$
where $[x]$ is the fractional part of $x$ for $x\geq0$ and $F$ is extended 
to $(-\infty,0)$ by $F(-x)=-F(x)$.\par
Its level sets are very similar to the ones of previous function. Below we show 
a picture of the level $N_0=f^{-1}(0)$, that has the same peculiarity of the
level set $M_0$ studied in previous section.\par
As before, this function in the range of energies $(-1,1)$ gives rise to genus-3 
surfaces embedded in $\tT$ with rank 3, so just 4 saddles of the foliation contribute to 
the topology of our system. All other saddles (if any) will be linked to a center 
and hence will be homotopic to 0 in the surface and easily eliminated from
the surface through a homotopy naturally generated by the center itself.\par
The analytical expression for the critical points for $E\in[-1,0]$
are the following:

$$\vbox{\halign{$#$\hfill&$#$\hfill&&$#$\hfill&$#$\hfill\cr
	x_1(a,b,E) &= \cases{\eqalign{
			\phantom{\frac{1}{2} - }\frac{a\sqrt(1+E)}{4\sqrt{1-a^2+b^2}}&\;,\;\;\;b^2-a^2\geq E\cr
			\phantom{\frac{1}{2} - }\frac{a\sqrt(1-E)}{4\sqrt{1+a^2-b^2}}&\;,\;\;\;b^2-a^2\leq E\cr
			}\cr}&\hskip1cm&x_4 &= 1 - x_1\cr
	\noalign{\vskip.5cm}
	y_1(a,b,E) &= \cases{\eqalign{
			\frac{1}{2} - \frac{b\sqrt(1+E)}{4\sqrt{1-a^2+b^2}}&\;,\;\;\;b^2-a^2\geq E\cr
			\frac{1}{2} - \frac{b\sqrt(1-E)}{4\sqrt{1+a^2-b^2}}&\;,\;\;\;b^2-a^2\leq E\cr
			}\cr}&&y_4 &= 1 - y_1\cr
	\noalign{\vskip.5cm}
	z_1(a,b,E) &= \cases{\eqalign{
			\frac{1}{2} - \frac{\sqrt(1+E)}{4\sqrt{1-a^2+b^2}}&\;,\;\;\;b^2-a^2\geq E\cr
			\frac{1}{2} - \frac{\sqrt(1-E)}{4\sqrt{1+a^2-b^2}}&\;,\;\;\;a^2-b^2\leq E\cr
			}\cr}&&z_4 &= 1 - z_1\cr
	\noalign{\vskip.5cm}
	x_2(a,b,E) &= \cases{\eqalign{
			\frac{1}{2} - \frac{a\sqrt(1-E)}{4\sqrt{1-a^2+b^2}}&\;,\;\;\;a^2-b^2\leq E\cr
			\frac{1}{2} - \frac{a\sqrt(1+E)}{4\sqrt{1+a^2-b^2}}&\;,\;\;\;a^2-b^2\geq E\cr
			}\cr}&&x_3 &= 1 - x_2\cr
	\noalign{\vskip.5cm}
	y_2(a,b,E) &= \cases{\eqalign{
			\phantom{\frac{1}{2} - }\frac{b\sqrt(1-E)}{4\sqrt{1-a^2+b^2}}&\;,\;\;\;a^2-b^2\leq E\cr
			\phantom{\frac{1}{2} - }\frac{b\sqrt(1+E)}{4\sqrt{1+a^2-b^2}}&\;,\;\;\;a^2-b^2\geq E\cr
			}\cr}&&y_3 &= 1 - y_2\cr
	\noalign{\vskip.5cm}
	z_2(a,b,E) &= \cases{\eqalign{
			\phantom{\frac{1}{2} - }\frac{\sqrt(1-E)}{4\sqrt{1-a^2+b^2}}&\;,\;\;\;a^2-b^2\leq E\cr
			\frac{1}{2} - \frac{\sqrt(1+E)}{4\sqrt{1+a^2-b^2}}&\;,\;\;\;a^2-b^2\geq E\cr
			}\cr}&&z_3 &= 1 - z_2\cr
}}$$

The expression of boundaries of all zones in this case is very simple:
for example the boundary of the zone labeled by $(0,0,1)$ at energy $E$ is 
the union of the segments of ellipse $(1+E)a^2+(3-E)b^2=1+E$ for $b\geq a$
and $(1+E)b^2+(3-E)a^2=1+E$ for $b\leq a$, and the boundary of the zone 
corresponding to $(1,1,1)$ is the union of the segments $8a-(3-E)a^2-(1+E)b^2=3-E$
for $b\geq a$ and $8b-(3-E)b^2-(1+E)a^2=3-E$ for $b\leq a$.\par
Using the same triples of integers used for boundaries in table \cosSk\ 
we have been able to find with a few modification the corresponding zones
for this function. As shown in table \parSk\ to this zones corresponds exactly 
the same homology classes of the previous picture, as we expected given 
the similarity between the two functions.\par
We analized numerically the stability zones in the square $[0,1]^2$ for the
same energies, finding the data reported in tables \parZ-\parN.
At every energy we included in the picture also the boundary of a few zones 
to show the very good agreement of numerical data with the analytical results.\par
Using the data found at resolution $1000$ (table XI) we evaluated again fractal 
dimension of the set of ergodic directions with the two methods used for
the trigonometric case, finding very similar results: the Box Counting method
gives us an estimate of $d\simeq1.77$ and from the growth rate of sizes of stability
zones we get $d/2\simeq.9$. Therefore the two different estimates are in very good
agreement also in this case and suggest a fractal dimension around $d=1.8$.

$$
\vbox{\halign{\hfill#\hfill\cr
\psset{unit=.8cm}
\pspicture(0,0)(9,4)
\psaxes[tickstyle=top,Dy=2,dy=.5]{->}(11,4)
\psdots*(1,.4)(2,.86)(3,1.34)(4,1.78)(5,2.23)(6,2.67)(7,3.1)(8,3.52)(9,3.86)
\rput(10.75,-.4){$n$}
\rput(-1.5,1.75){$\log_2(N_n)$}
\endpspicture\cr
\noalign{\vskip .9cm}
\caption\cylinders{plot of the $\log$ in base 2 of number of squares needed to 
cover the fractal with squares of size $2^{2n}$ versus the size scale $n$.}\cr
\noalign{\vskip 1.5cm}
\pspicture(0,0)(9,4)
\psaxes[tickstyle=top,Dy=2,dy=.5,Dx=2,dx=1]{->}(9.75,4)
\psdots*(.5,0)(1,0)(1.5,0)(2,.25)(2.5,.25)(3,.75)(3.5,.65)(4,1.04)(4.5,1.18)(5,1.43)
(5.5,1.63)(6,1.84)(6.5,2.08)(7,2.3)(7.5,2.58)(8,2.92)(8.5,3.57)
\rput(9.45,-.4){$n$}
\rput(-1.5,1.75){$\log_2(N_n)$}
\endpspicture\cr
\noalign{\vskip .9cm}
\caption\cylinders{plot of the $\log$ in base 2 of number of number of zones of area 
between $2^{-n}$ and $2^{-n-1}$ versus the size scale $n$.}\cr
}}
$$

\section\void{Conclusions}

We produced a C++ library that implements all functions needed to analyze
numerically the topological behaviour of foliations induced on a periodic 
surface of genus 3 by a constant 1-form. 
This problem is equivalent in physics to the
behaviour of magnetoresistance under a strong magnetic field.\par
We checked numerically our code on two ``toy functions'' that produce
genus 3 surfaces embedded in $\tT$ with rank 3 and verified its 
correctness comparing numerical data with the analytical data that was 
possible to get for the two simple functions chosen, finding a very good 
agreement between the two.\par
Our next move will be to apply this machinery to concrete Fermi surfaces 
of normal metals, that have genus 4 in the easiest cases.

\Acknowledgments

The author gratefully wants to thank his advisor S.P. Novikov for his 
numerous advices and for several helpful discussions about the subject.
The author also acknowledge many fruitful discussions with I.A. Dynnikov,
A. Giacobbe, B. Hunt, D.J. Patil and K. Snitz.\par
The author also thanks Indam for financial support and the IPST, the Cagliari (Italy)
section of INFN and the High School ``L.B. Alberti'' in Cagliari for providing
the several Alpha Digital and Pentium II Linux Machines on which all numerical 
simulations have been run.

\theBibliography

\bibitem[ASY96]\
K.T. Alligood, T.D. Sauer \& J.A. Yorke, {\sl Chaos: an introduction to dynamical systems}, 1996, Springer Verlag
\bibitem[DL99]\ 
R. De Leo, {\sl Existence and measure of ergodic leaves in Novikov problem on the semiclassical motion of an electron}, 
Usp. Mat. Nauk (RMS), 54:6 (1999), {\bf math-ph/0005031}
\bibitem[Dyn93a]\ 
I. Dynnikov, {\sl Proof of S.P. Novikov's conjecture on the semiclassical motion of an electron}, 
Mat. Zametki 53:5 (1993), 57-68
\bibitem[Dyn97]\ 
I. Dynnikov, {\sl Semiclassical motion of the electron. A proof of the Novikov conjecture in general position
and counterexamples}, AMS Transl., 179 (1997), 45-73
\bibitem[Fal97]\
K. Falconer, {\sl Techniques in fractal geometry}, 1997, Wiley
\bibitem[Nov82]\ 
S.P. Novikov, {\sl Hamiltonian formalism and a multivalued analog of Morse theory}, 
Usp. Mat. Nauk (RMS), 37:5 (1982), 3-49
\bibitem[NM98]\ 
S.P. Novikov and A.Ya. Maltsev, {\sl Topological phenomena in normal metals}, Usp. Fiz. Nauk, 41:3, (1998), 231-239, {\bf cond-mat/9709007 }
\bibitem[Zor84]\ 
A.V. Zorich, {\sl A problem of Novikov on the semiclassical motion of electrons in a uniform almost rational magnetic field},
Usp. Mat. Nauk (RMS), 39:5 (1984), 235-236

\vfill\eject

$$
\vbox{\tabskip=20pt\halign{\hfill#\hfill\cr
\pspicture(0,0)(9,20)
\rput(4.5,10){\vbox{\halign{\hfill#\hfill&\hfill#\hfill&\hfill#\hfill\cr
	\epsfxsize=5cm\epsfbox{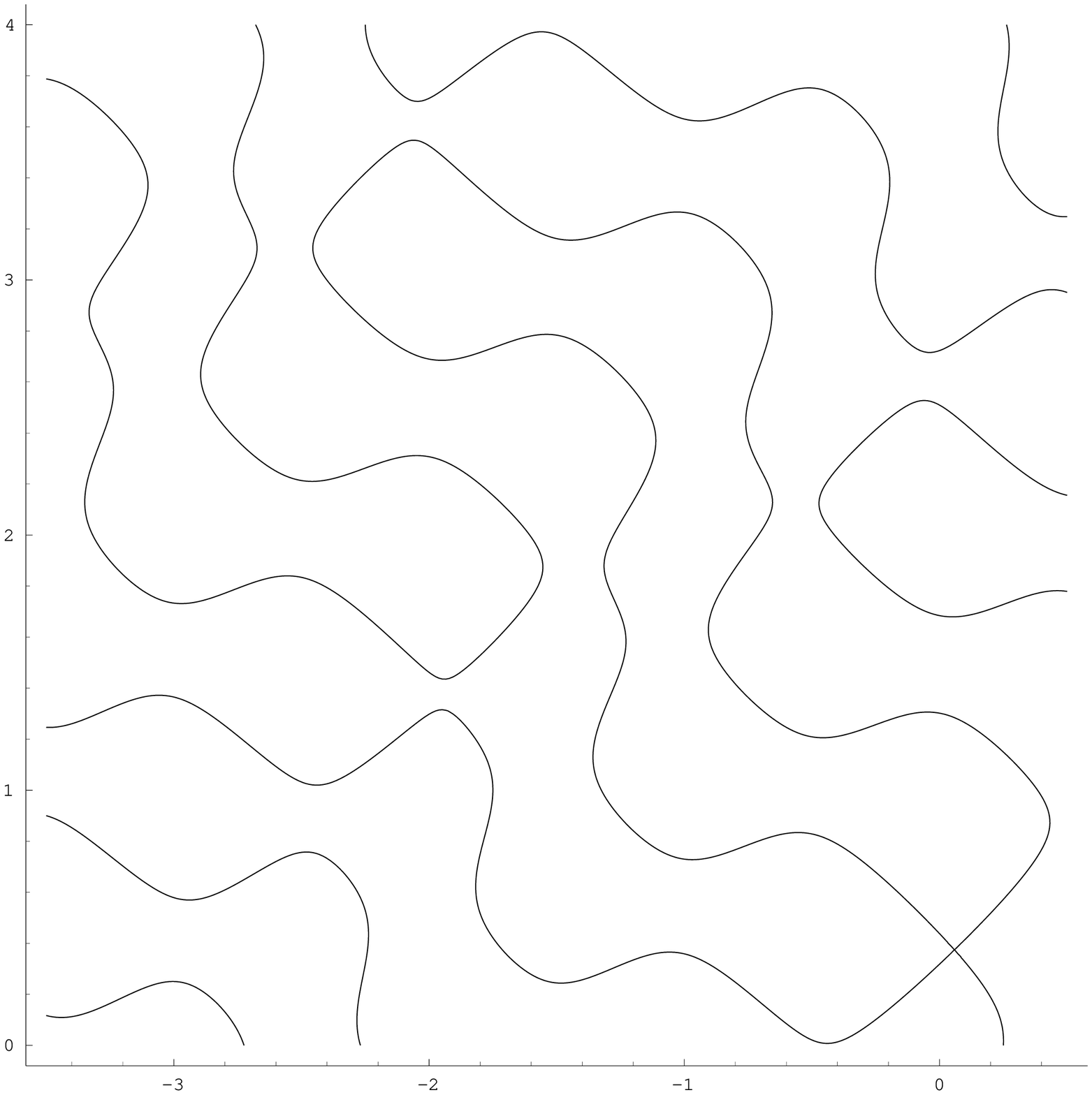}
&
	\epsfxsize=5cm\epsfbox{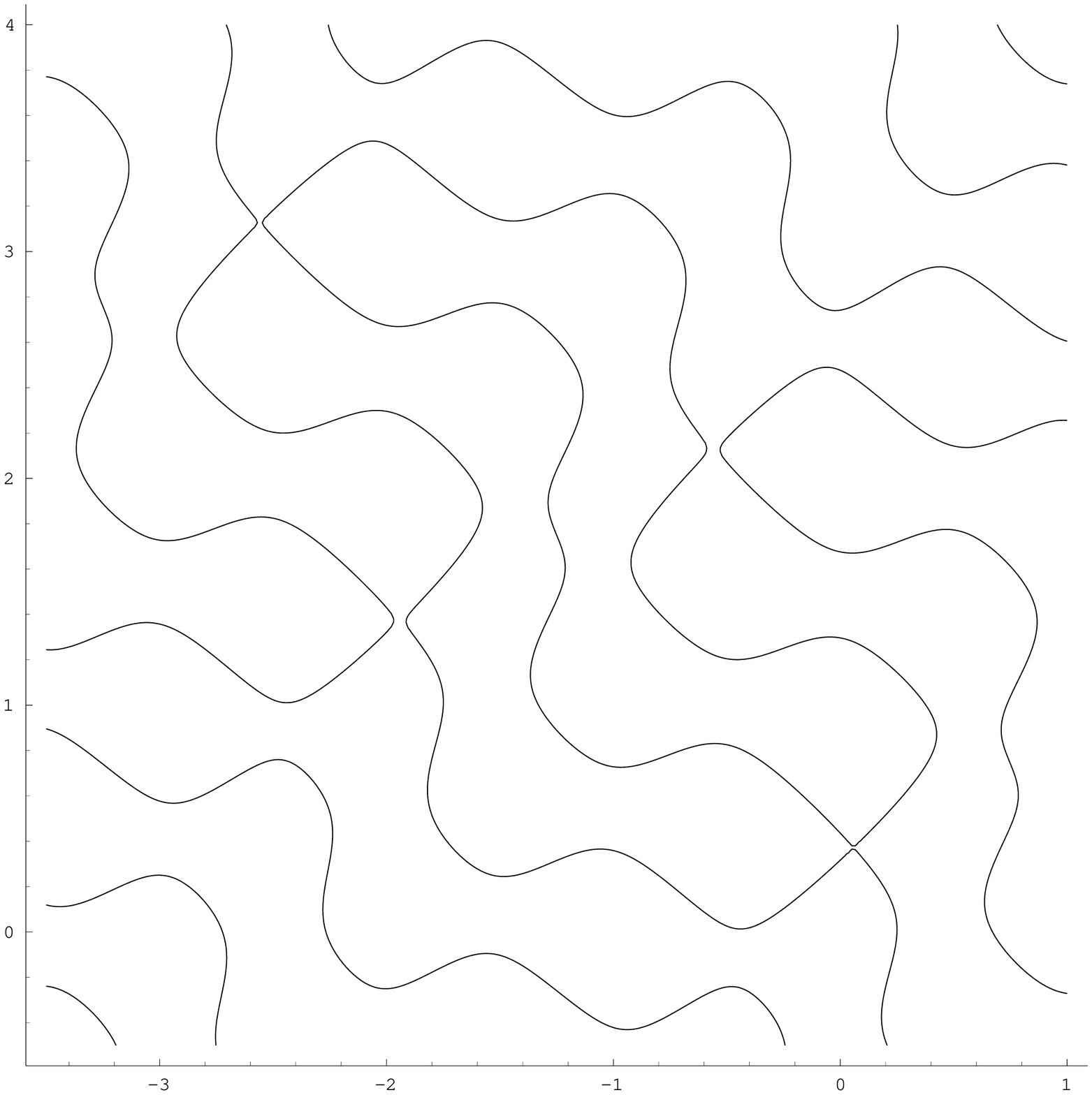}
&
	\epsfxsize=5cm\epsfbox{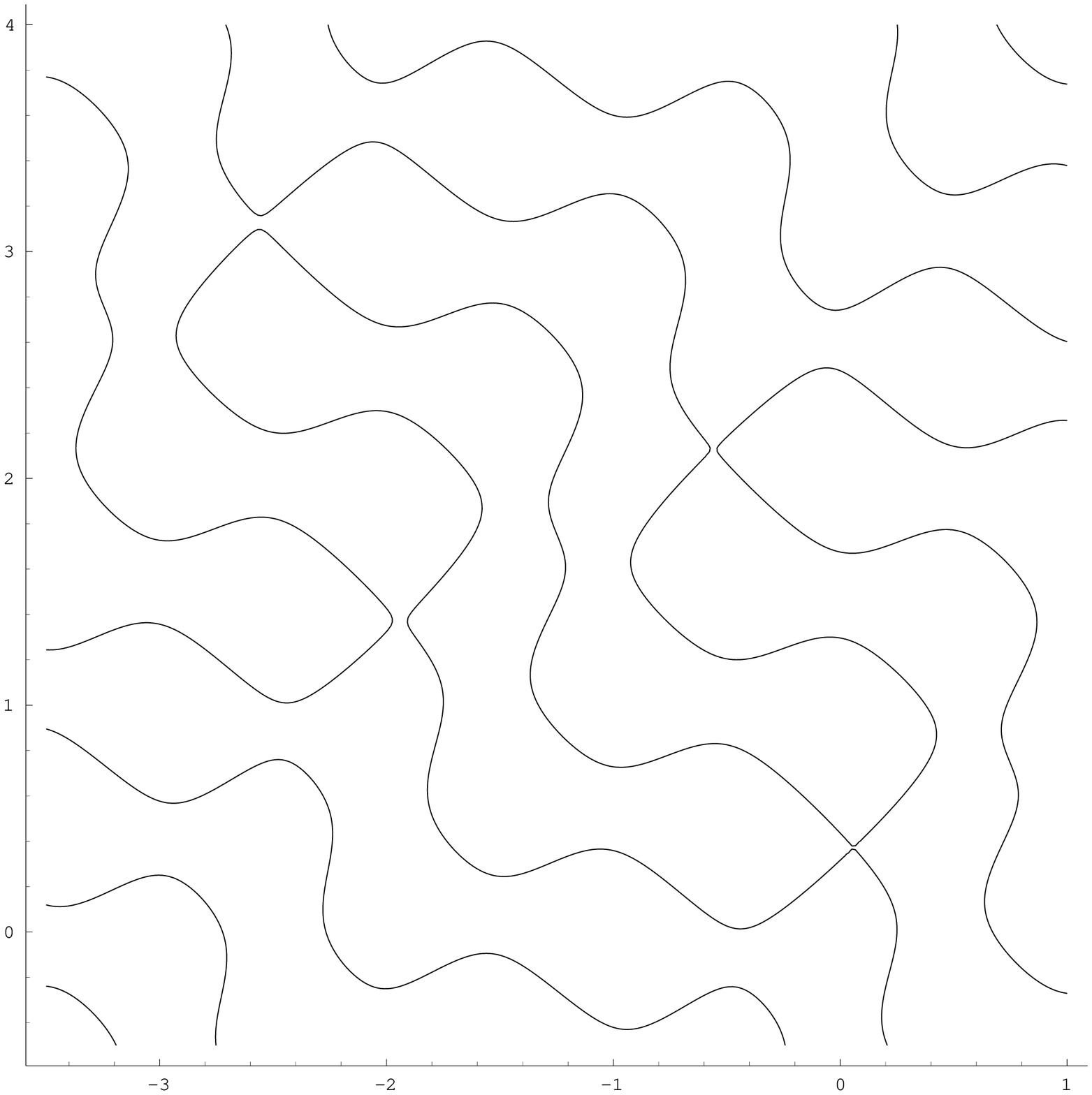}
\cr
	\noalign{\vskip5pt}
	{\bf 1.}\ \ $H = (.37,.73,1)$
& 
	{\bf 2.}\ \ $H \simeq (.37,.742,1)$
& 
	{\bf 3.}\ \ $H = (.37,.743,1)$
\cr 
	\noalign{\vskip20pt}
	\epsfxsize=5cm\epsfbox{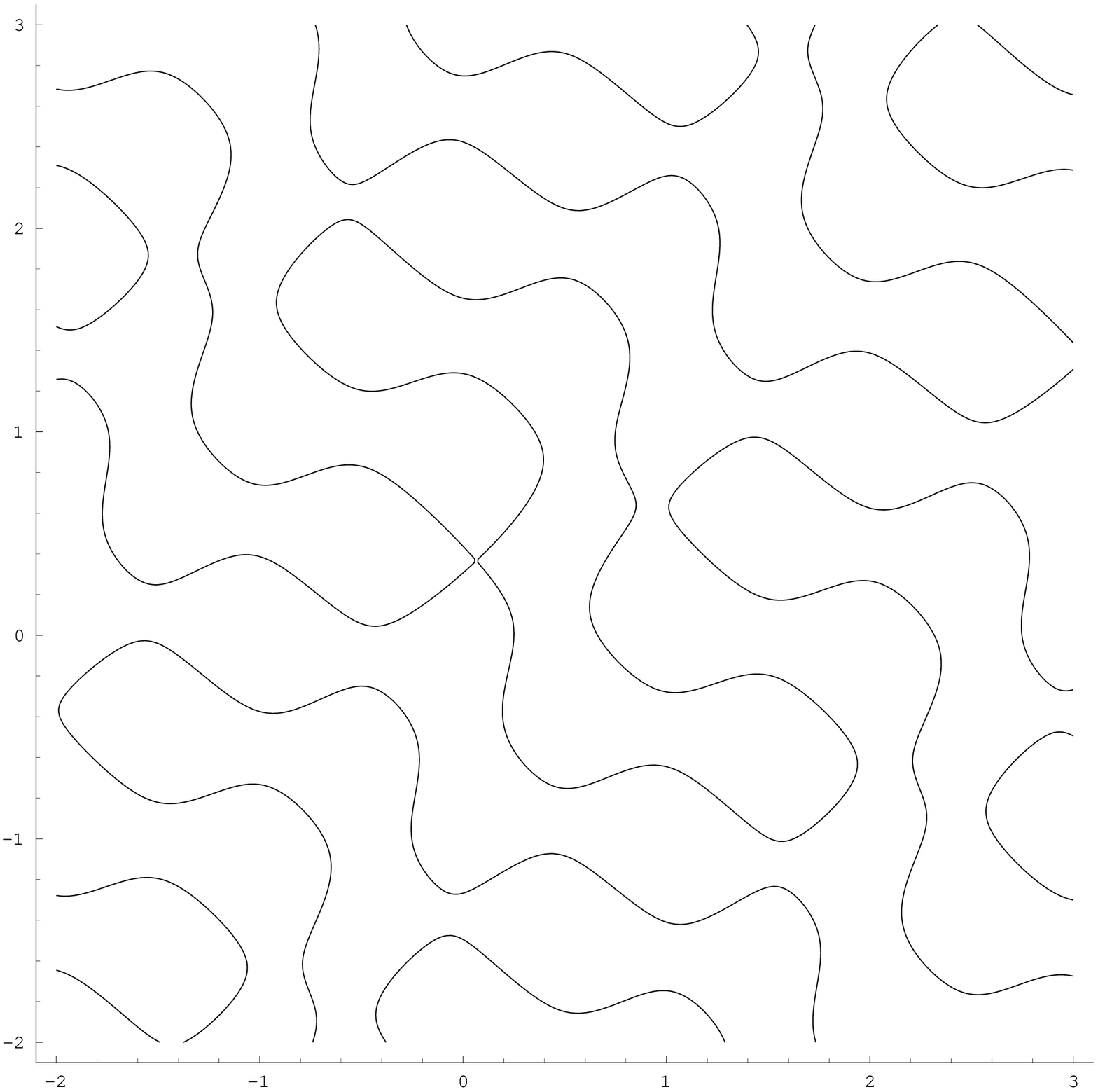}
&
	\epsfxsize=5cm\epsfbox{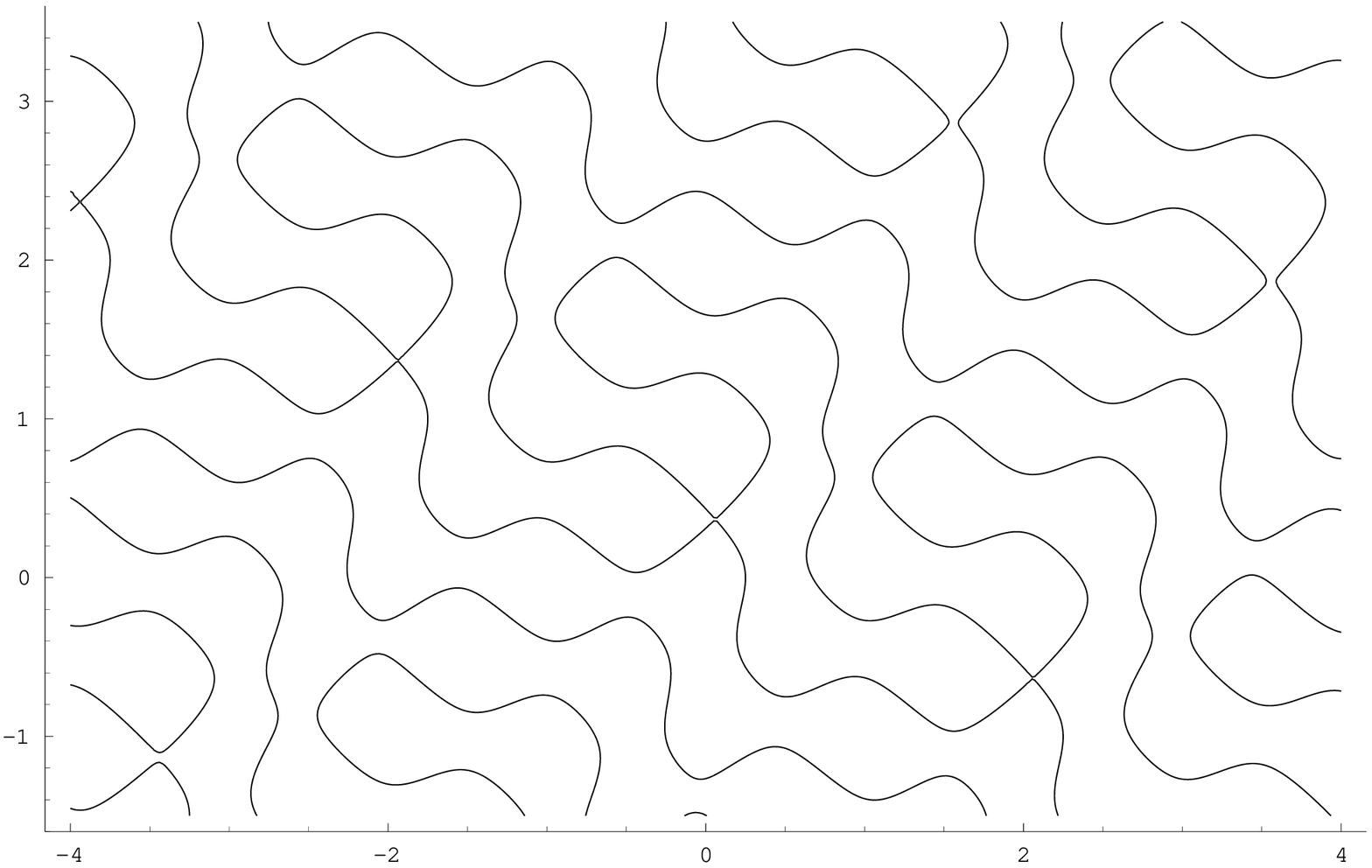}
&
	\epsfxsize=5cm\epsfbox{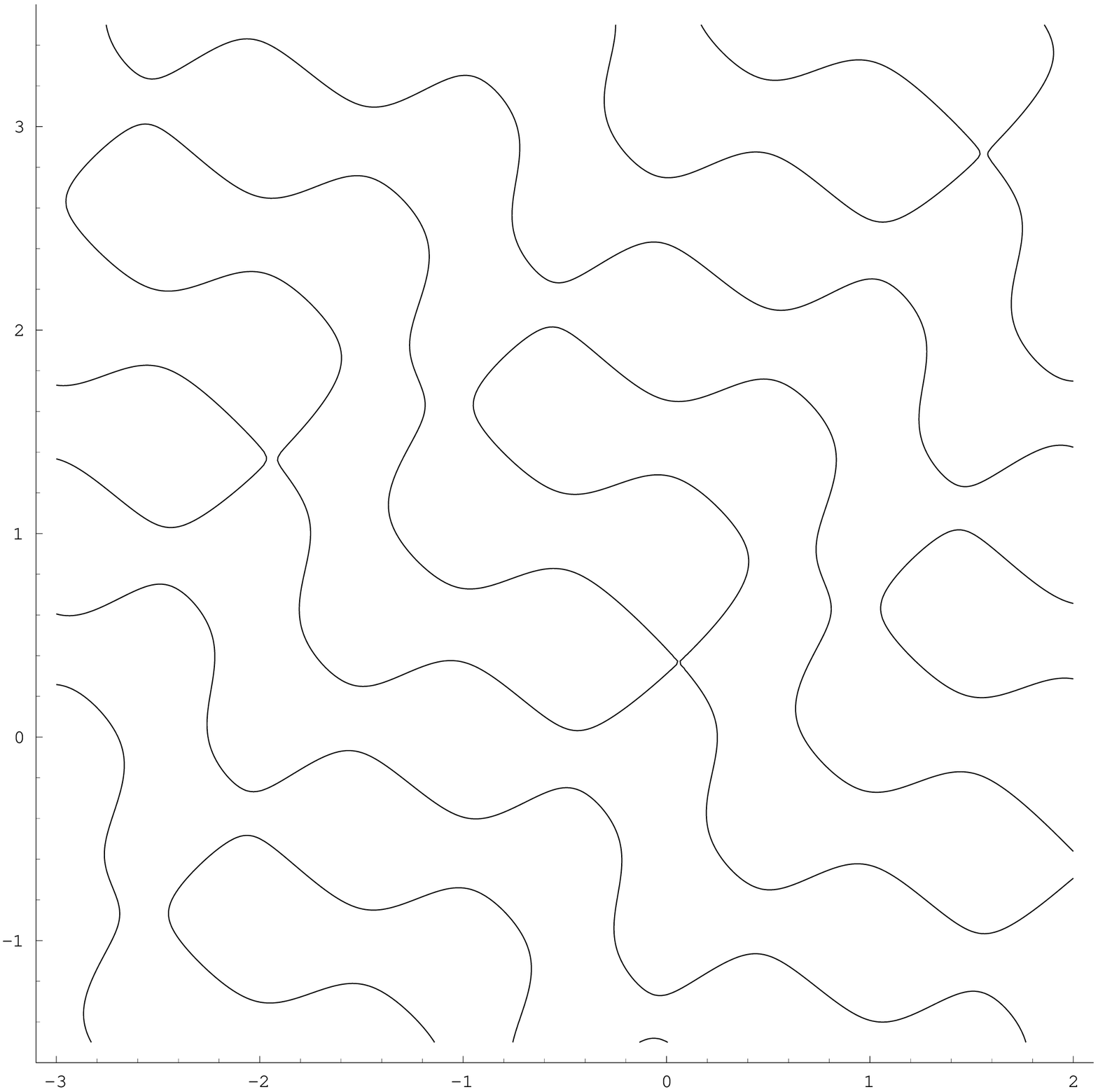}\cr
	\noalign{\vskip5pt}
	{\bf 4.}\ \ $H = (.41,.77,1)$
& 
	{\bf 5.}\ \ $H = (.385,.77,1)$
& 
	{\bf 6.}\ \ $H = (.384,.77,1)$
\cr 
	\noalign{\vskip20pt}
	\epsfxsize=5cm\epsfbox{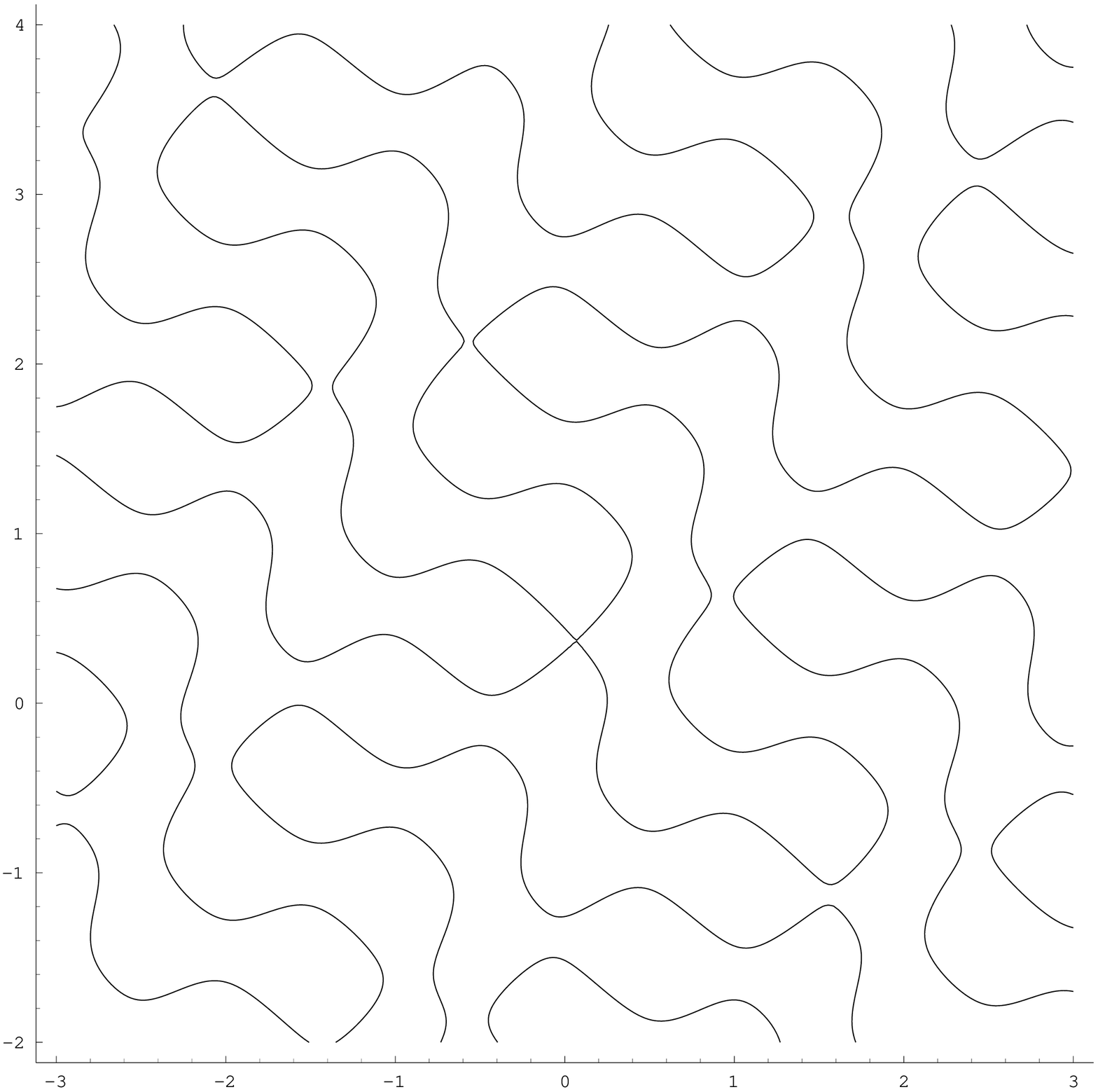}
&
	\epsfxsize=5cm\epsfbox{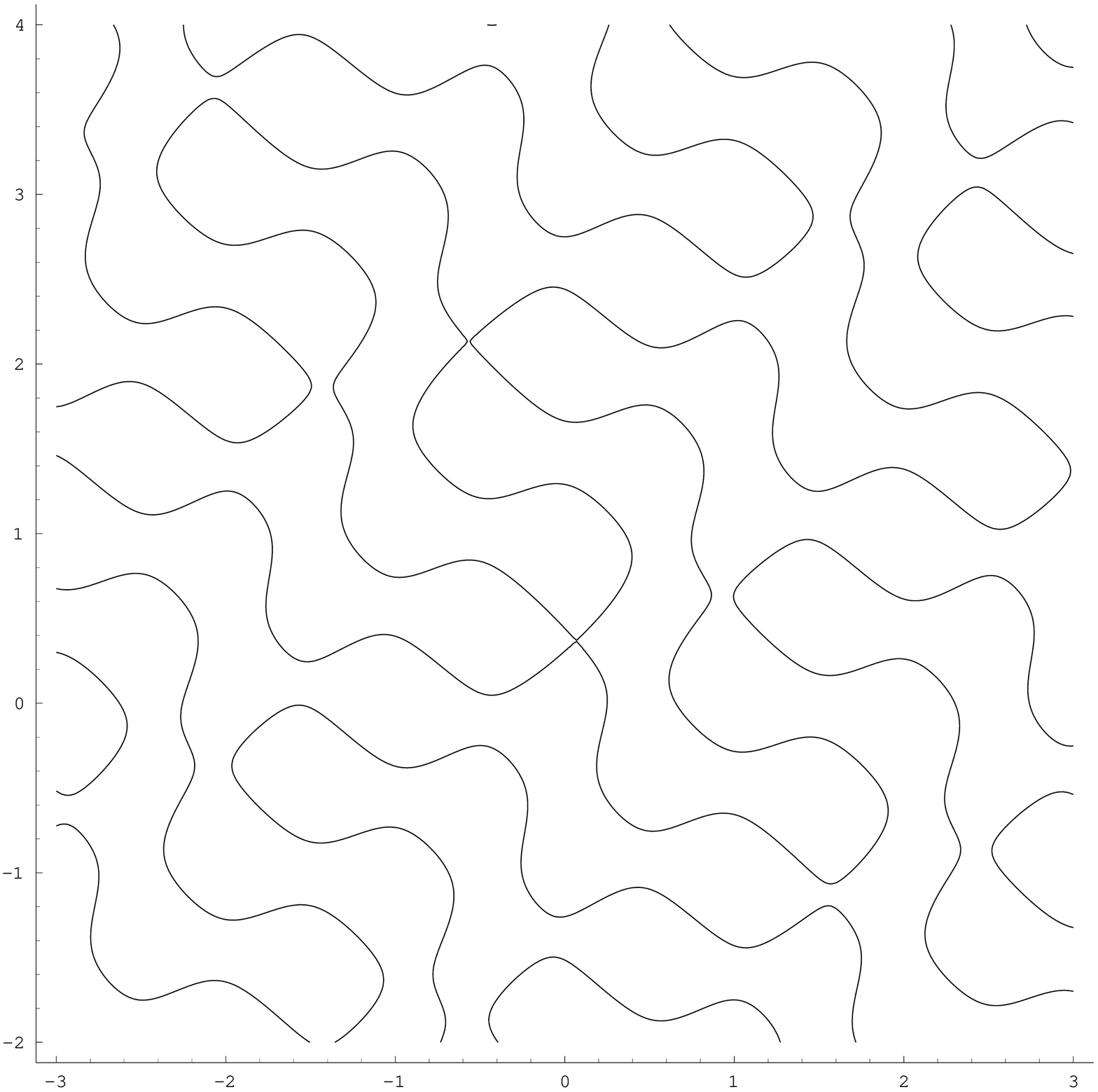}
&
	\epsfxsize=5cm\epsfbox{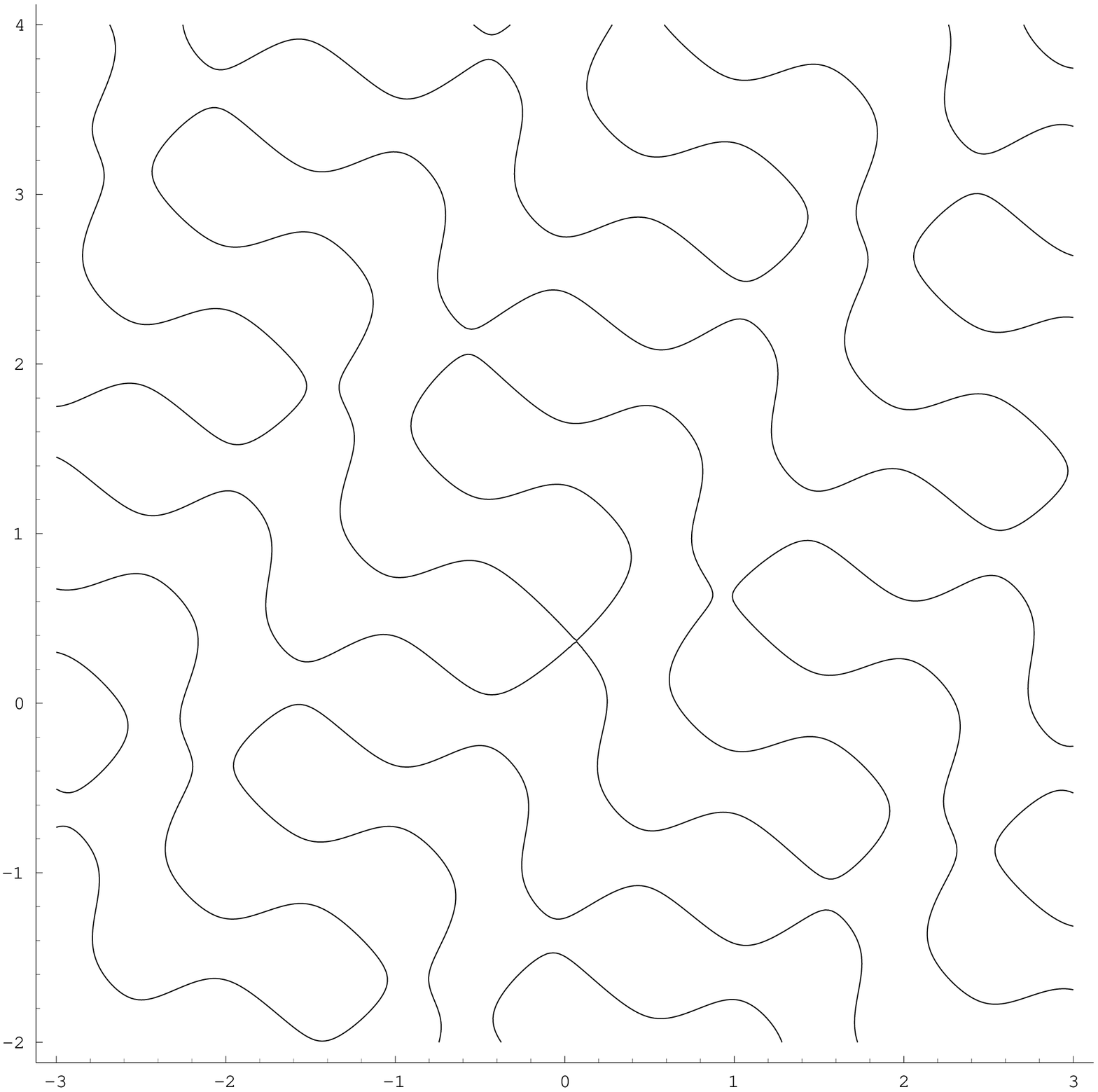}\cr
	\noalign{\vskip5pt}
	{\bf 7.}\ \ $H = (.42,.76,1)$
& 
	{\bf 8.}\ \ $H \simeq (.42,.761,1)$
& 
	{\bf 9.}\ \ $H = (.42,.77,1)$
\cr 
}}}
\endpspicture\cr
\tableCaption\void{to illustrate the phenomenon of the change of cylinder type 
inside a stability zone we show what happens in case of the zone $(2,4,5)$ shown
in figure \zoneTFF. 
\par\noindent
{\bf 1-3}: in the first row we move the direction of $H$ from subzone \rmU to subzone
\rmT. In the central picture we reach the boundary between the subzones, in which point 
$p_1$ has a saddle connection with point $p_2+(-3,3,-2)$.
\par\noindent
{\bf 4-6}: in the second row we move $H$ from subzone \rmT to subzone \rmD. This
time on boundary $p_1$ has a saddle connection with a copy of itself separated by
a 1-rational vector $(2,-1,0)$.  
\par\noindent
{\bf 7-9}: in last row we move $H$ from subzone \rmU to subzone \rmD. At the 
boundary $p_1$ has a saddle connection with $p_2+(1,-2,1)$.
}\cr
}}
$$

\vfill\eject

\null
$$
\vbox{\halign{\hfill#\hfill\cr
\pspicture(0,0)(16,16)
\rput(8,9.5){\epsfxsize=27cm\epsfbox{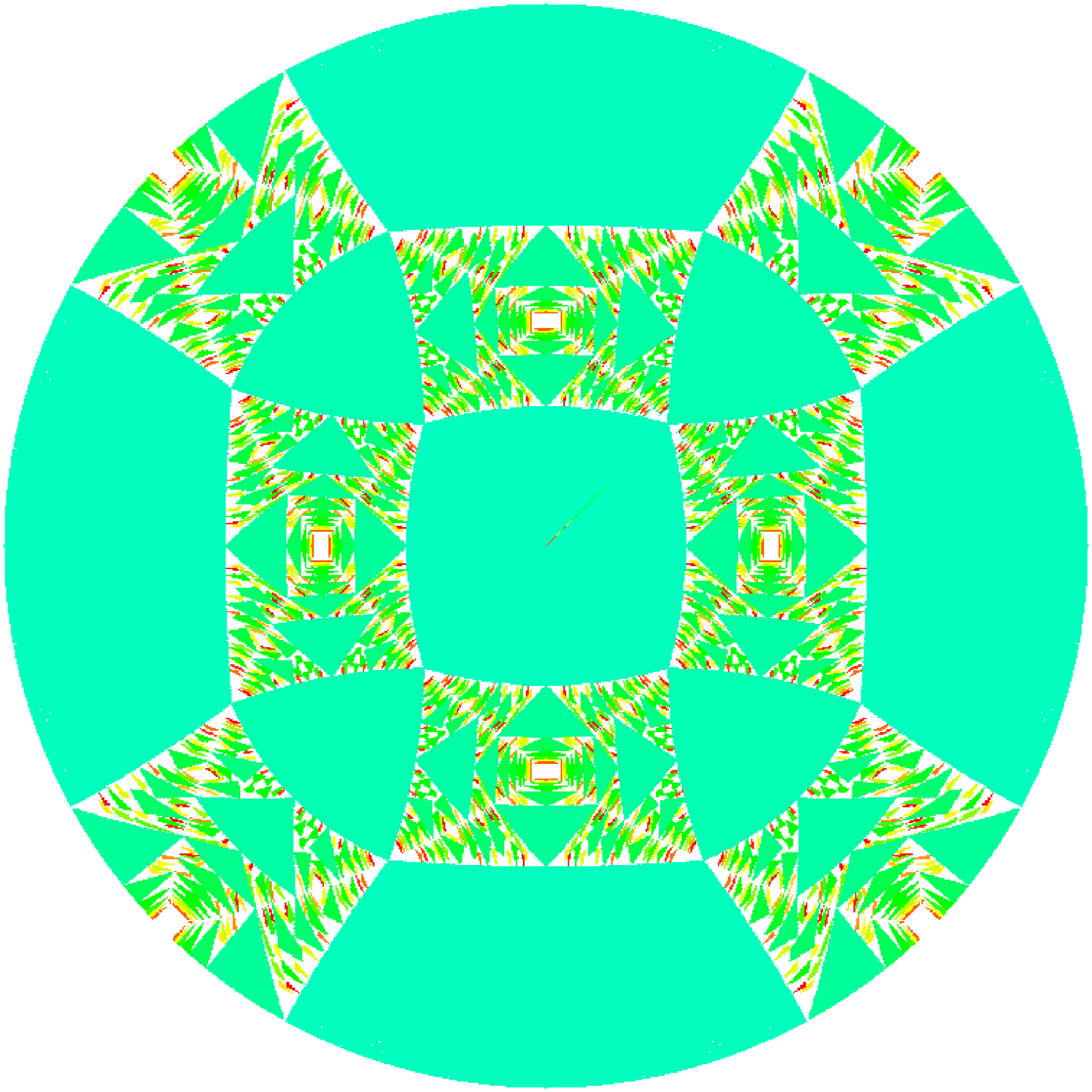}}
\endpspicture\cr
\tableCaption\novikovFractalDisc{the fractal picture in the square $[0,1]^2$
obtained at a resolution $N=10^3$. Of the $\sim3\cdot10^4$ zones found just the
ones with at least 10 points ($\sim1000$) are shown. The square has been 
obtained just symmetrizing the triangular picture obtained. To get
this picture we used 5 Linux machines with Pentium II CPUs for $\sim3$ weeks.
It is possible to get the homology class corresponding to the biggest zones
comparing this picture with next one and with the table included in next
page.
From these data has been extrapolated a fractal dimension of $d\simeq1.77$ for the
set of ``ergodic'' directions.}\cr
}}
$$

\vfill\eject

\null
$$
\vbox{\halign{\hfill#\hfill\cr
\pspicture(0,0)(16,16)
\rput(8,8.5){\epsfxsize=17cm\epsfbox{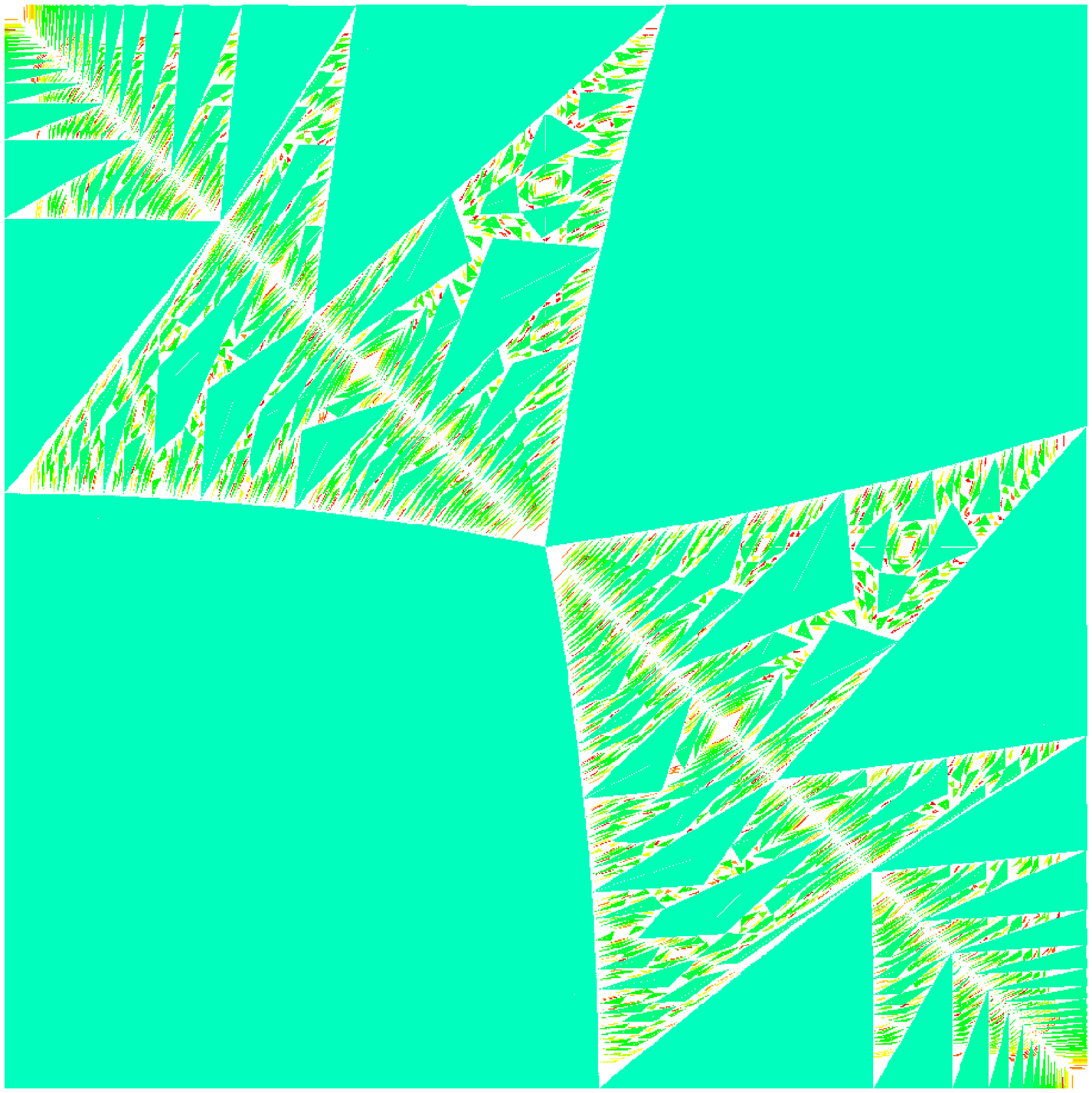}}
\endpspicture\cr
\tableCaption\novikovFractal{the fractal picture in the square $[0,1]^2$
obtained at a resolution $N=10^3$. Of the $\sim3\cdot10^4$ zones found just the
ones with at least 10 points ($\sim1000$) are shown. The square has been 
obtained just symmetrizing the triangular picture obtained. To get
this picture we used 5 Linux machines with Pentium II CPUs for $\sim3$ weeks.
It is possible to get the homology class corresponding to the biggest zones
comparing this picture with next one and with the table included in next
page.
From these data has been extrapolated a fractal dimension of $d\simeq1.77$ for the
set of ``ergodic'' directions.}\cr
}}
$$

\vfill\eject

\null
\vskip -1.cm
$$
\vbox{\halign{\hfill#\hfill\cr
\pspicture(0,0)(16,16)
\rput(8,8){\epsfxsize=15cm\epsfbox{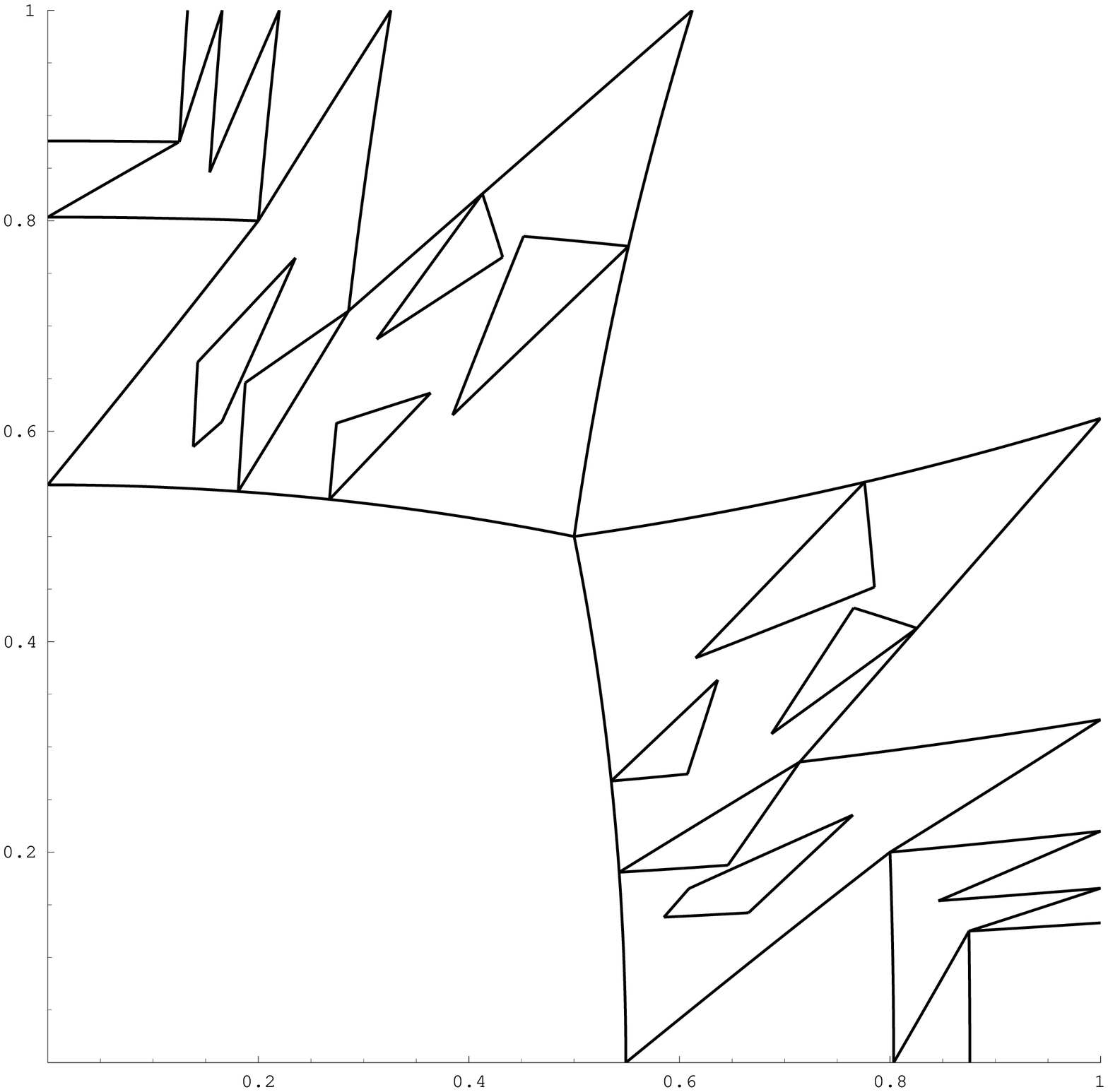}}
\rput(4.93,4.83){\rm(0,0,1)}
\rput(12.22,12.12){\rm(1,1,1)}
\rput(6.96,14.09){\rm(1,2,2)}
\rput(14.19,6.86){\rm(2,1,2)}
\rput(2.13,11.46){\rm(0,1,2)}  
\rput(11.56,2.03){\rm(1,0,2)}
\rput(4.67,14.55){\sixrm(1,3,3)}
\rput(14.65,4.57){\sixrm(3,1,3)}
\rput(11.6,7.67){\sixrm(3,2,4)}
\rput(7.76,11.49){\sixrm(2,3,4)}
\rput(3.55,10.48){\fiverm(1,4,6)}
\rput(4.2,10.2){\sixrm(1,3,5)}
\rput(10.05,3.95){\sixrm(3,1,5)}
\rput(10.69,3.49){\sixrm(4,1,6)}
\rput(1.74,13.2){\sixrm(0,2,3)}
\rput(13.2,1.34){\sixrm(2,0,3)}
\rput(3.9,15.3){\sixrm(1,4,4)}
\rput(6.88,11.98){\sixrm(2,4,5)}
\rput(12.08,6.58){\sixrm(4,2,5)}
\rput(14.84,3.59){\sixrm(4,1,4)}
\rput(3.25,15.5){\sixrm(1,5,5)}
\rput(14.95,3.03){\sixrm(5,1,5)}
\rput(5.47,9.58){\sixrm(1,2,4)}
\rput(9.47,5.1){\sixrm(2,1,4)}
\endpspicture\cr
\tableCaption\cosSk{boundaries of a few stability zones together with
their homology class. All these boundaries have been obtained with the
technique explained in section \cosSec, while the associated homology
class has been obtained numerically, except in the trivial case of 
(0,0,1). Below are listed the zones with biggest size and their area,
form the data found at $N=1000$.}\cr
}}
$$

$$
\vbox{\tabskip=0pt \offinterlineskip
\def\tablerule{\noalign{\hrule}}
\halign{
  \strut#&\vrule#\tabskip=1em plus2em&
  \hfil$#$\hfil&\vrule#&\hfil$#$\hfil&\vrule#&
  \hfil$#$\hfil&\vrule#&\hfil$#$\hfil&\vrule#&
  \hfil$#$\hfil&\vrule#&\hfil$#$&\vrule#\tabskip=0pt\cr
\tablerule
&&\omit\hidewidth Hom Class\hidewidth&&\omit\hidewidth Area\hidewidth
&&\omit\hidewidth Hom Class\hidewidth&&\omit\hidewidth Area\hidewidth
&&\omit\hidewidth Hom Class\hidewidth&&\omit\hidewidth Area\hidewidth&\cr
\tablerule
&&(0,0,1)&&(2.83\pm.02)10^{-1}&&(2,4,5)&&(8.6\pm.6)10^{-3}&&(1,6,6)&&(2.0\pm.1)10^{-3}&\cr
\tablerule
&&(1,1,1)&&(2.03\pm.01)10^{-1}&&(1,4,4)&&(8.3\pm.3)10^{-3}&&(4,5,8)&&(2.0\pm.4)10^{-3}&\cr
\tablerule
&&(1,2,2)&&(8.2\pm.2)10^{-2}&&(1,2,4)&&(6.2\pm.5)10^{-3}&&(5,8,10)&&(1.9\pm.4)10^{-3}&\cr
\tablerule
&&(0,1,2)&&(5.1\pm.1)10^{-2}&&(3,4,6)&&(4.7\pm.5)10^{-3}&&(4,6,9)&&(1.8\pm.3)10^{-3}&\cr
\tablerule
&&(1,3,3)&&(2.1\pm.1)10^{-2}&&(1,5,5)&&(4.1\pm.2)10^{-3}&&(1,6,10)&&(1.7\pm.1)10^{-3}&\cr
\tablerule
&&(2,3,4)&&(1.7\pm.1)10^{-2}&&(2,5,8)&&(4.1\pm.4)10^{-3}&&(5,9,11)&&(1.6\pm.2)10^{-3}&\cr
\tablerule
&&(1,3,5)&&(9.6\pm.5)10^{-3}&&(4,7,8)&&(3.0\pm.3)10^{-3}&&(4,6,7)&&(1.5\pm.2)10^{-3}&\cr
\tablerule
&&(1,4,6)&&(9.6\pm.5)10^{-3}&&(0,3,4)&&(2.9\pm.4)10^{-3}&&(2,3,6)&&(1.5\pm.4)10^{-3}&\cr
\tablerule
&&(0,2,3)&&(9.0\pm.6)10^{-3}&&(3,5,7)&&(2.7\pm.3)10^{-3}&&(3,5,9)&&(1.5\pm.4)10^{-3}&\cr
\tablerule
}}$$

\vfill\eject

\null
\vskip 1.cm
$$
\vbox{\halign{\hfill#\hfill\cr
\pspicture(0,0)(16,16)
\rput(8,8){\epsfxsize=15cm\epsfbox{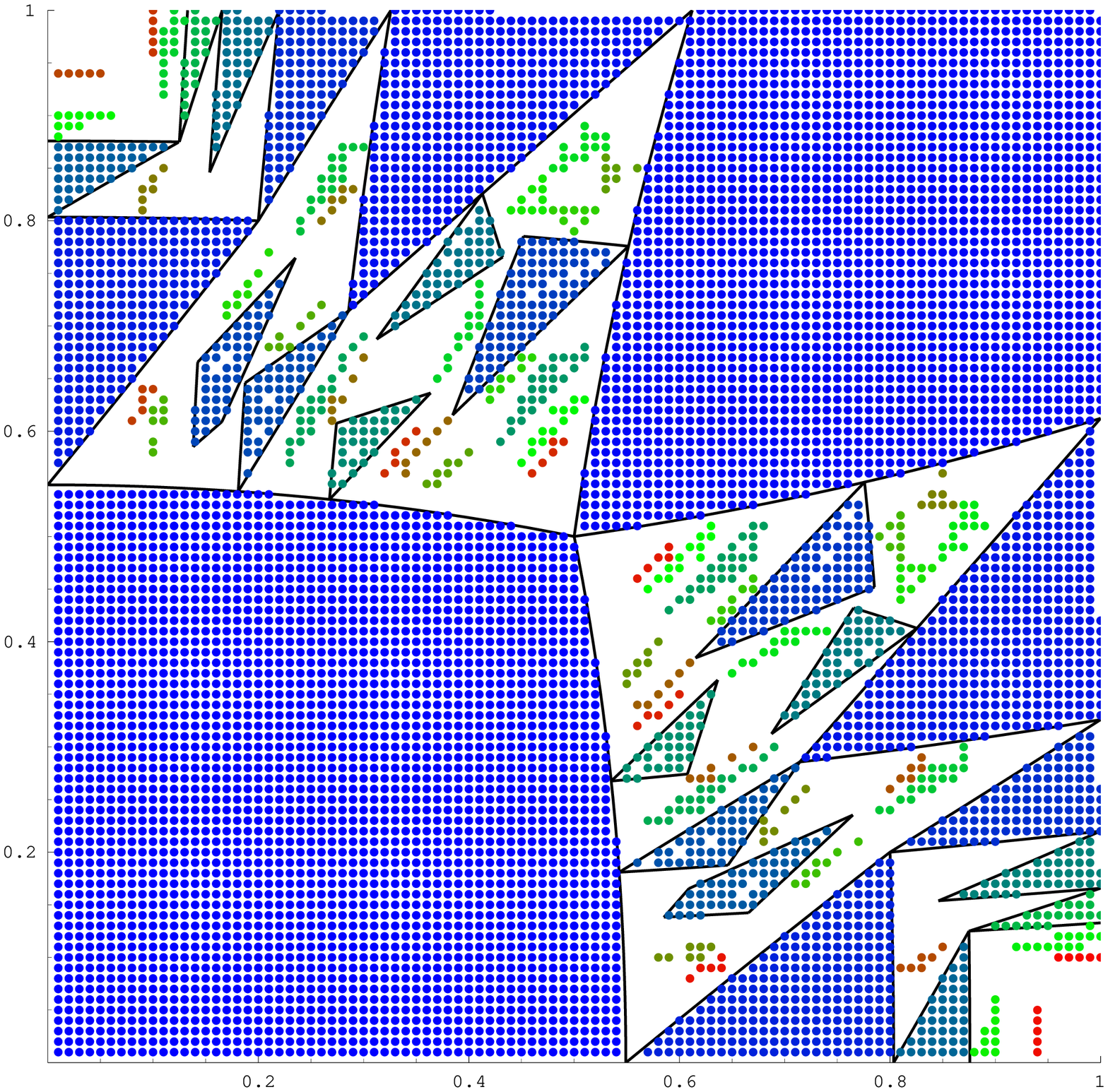}}
\endpspicture\cr
\tableCaption\cosZ{map of the stability zones at energy 0 in the square 
$[0,1]^2$ sampled at a resolution $N=100$. Nearly 700 zones are found at 
this resolution; in the above picture we show just the 74 that contain
at least 5 points. The boundary found analytically is also shown for a few
zones to show the perfect agreement with the numerical results.}\cr
}}
$$

\vfill\eject

\null
\vskip 1.cm
$$
\vbox{\halign{\hfill#\hfill\cr
\pspicture(0,0)(16,16)
\rput(8,8){\epsfxsize=15cm\epsfbox{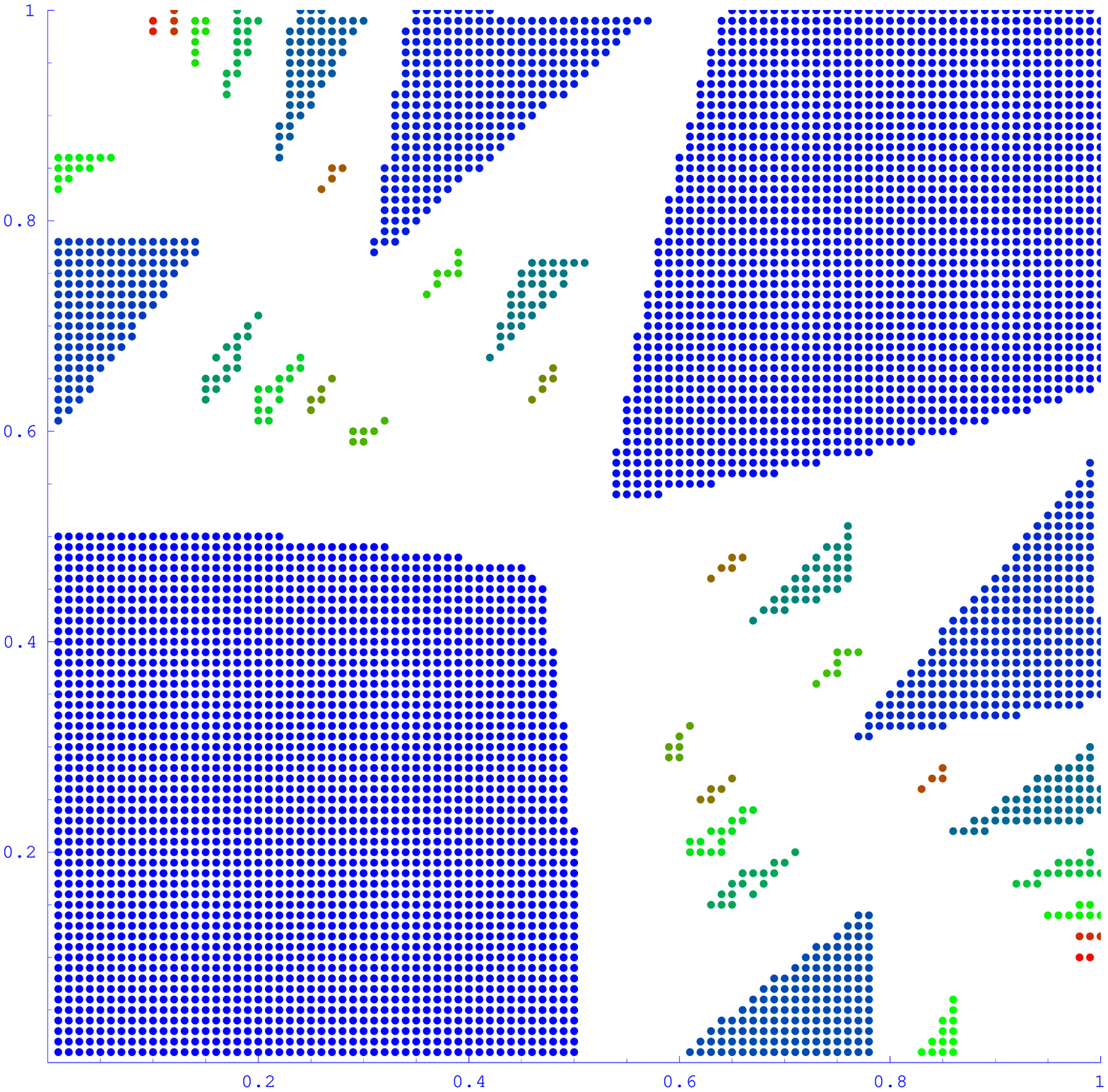}}
\endpspicture\cr
\tableCaption\cosO{map of the stability zones at energy $E=-.1$ in the square 
$[0,1]^2$ sampled at a resolution $N=100$. Just 48 zones are left at this energy,
and here we plotted just the 34 ones with more than 1 point.}\cr
}}
$$

\vfill\eject

\null
\vskip 1.cm
$$
\vbox{\halign{\hfill#\hfill\cr
\pspicture(0,0)(16,16)
\rput(8,8){\epsfxsize=15cm\epsfbox{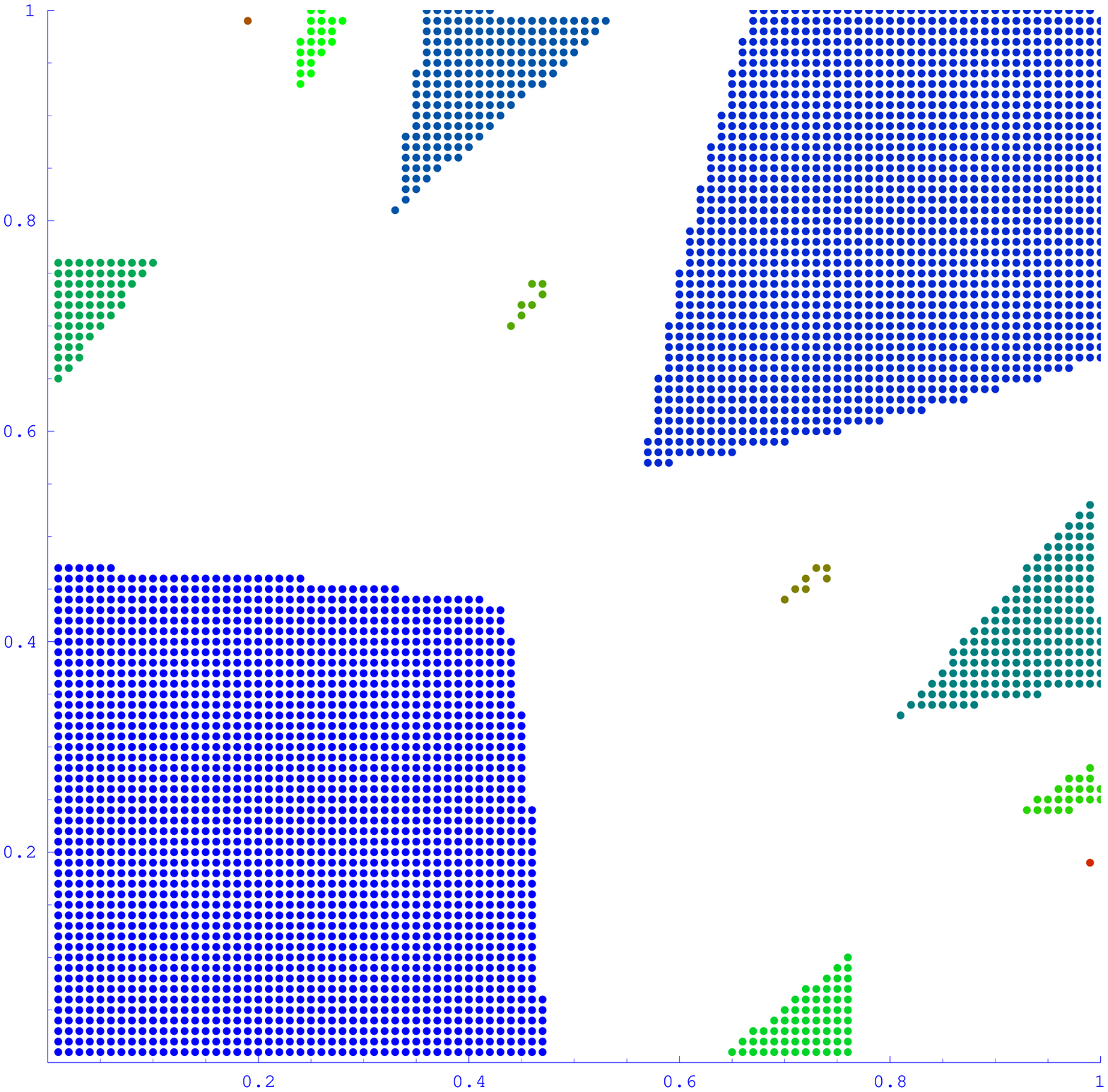}}
\endpspicture\cr
\tableCaption\cosTw{map of the stability zones at energy $E=-.2$ in the square 
$[0,1]^2$ sampled at a resolution $N=100$. Just 12 zones are left, and
we show all of them in the picture above.}\cr
}}
$$

\vfill\eject

\null
\vskip 1.cm
$$
\vbox{\halign{\hfill#\hfill\cr
\pspicture(0,0)(16,16)
\rput(8,8){\epsfxsize=15cm\epsfbox{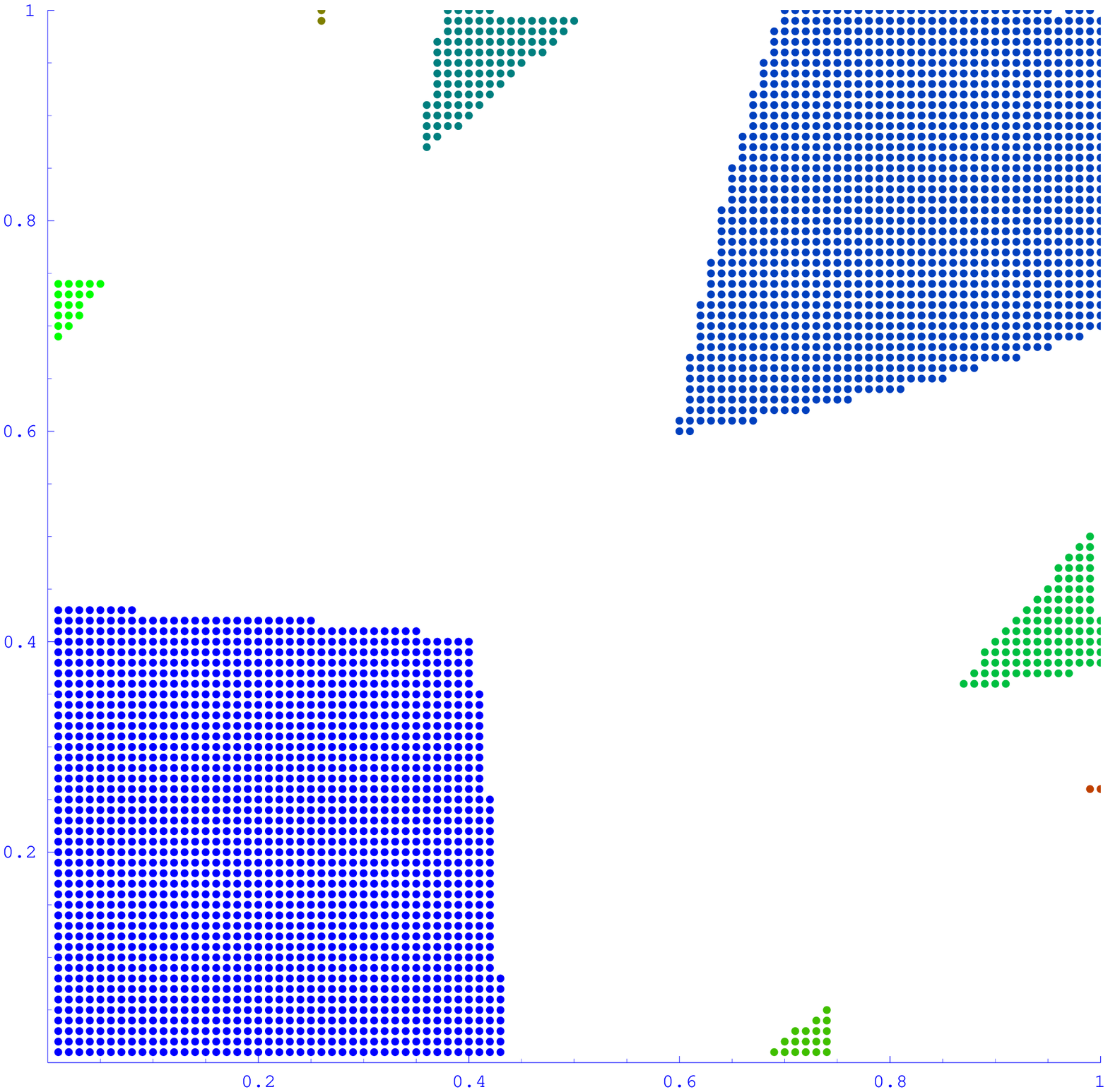}}
\endpspicture\cr
\tableCaption\cosTh{map of the stability zones at energy $E=-.3$ in the square 
$[0,1]^2$ sampled at a resolution $N=100$. All 8 zones found are shown.}\cr
}}
$$

\vfill\eject

\null
\vskip 1.cm
$$
\vbox{\halign{\hfill#\hfill\cr
\pspicture(0,0)(16,16)
\rput(8,8){\epsfxsize=15cm\epsfbox{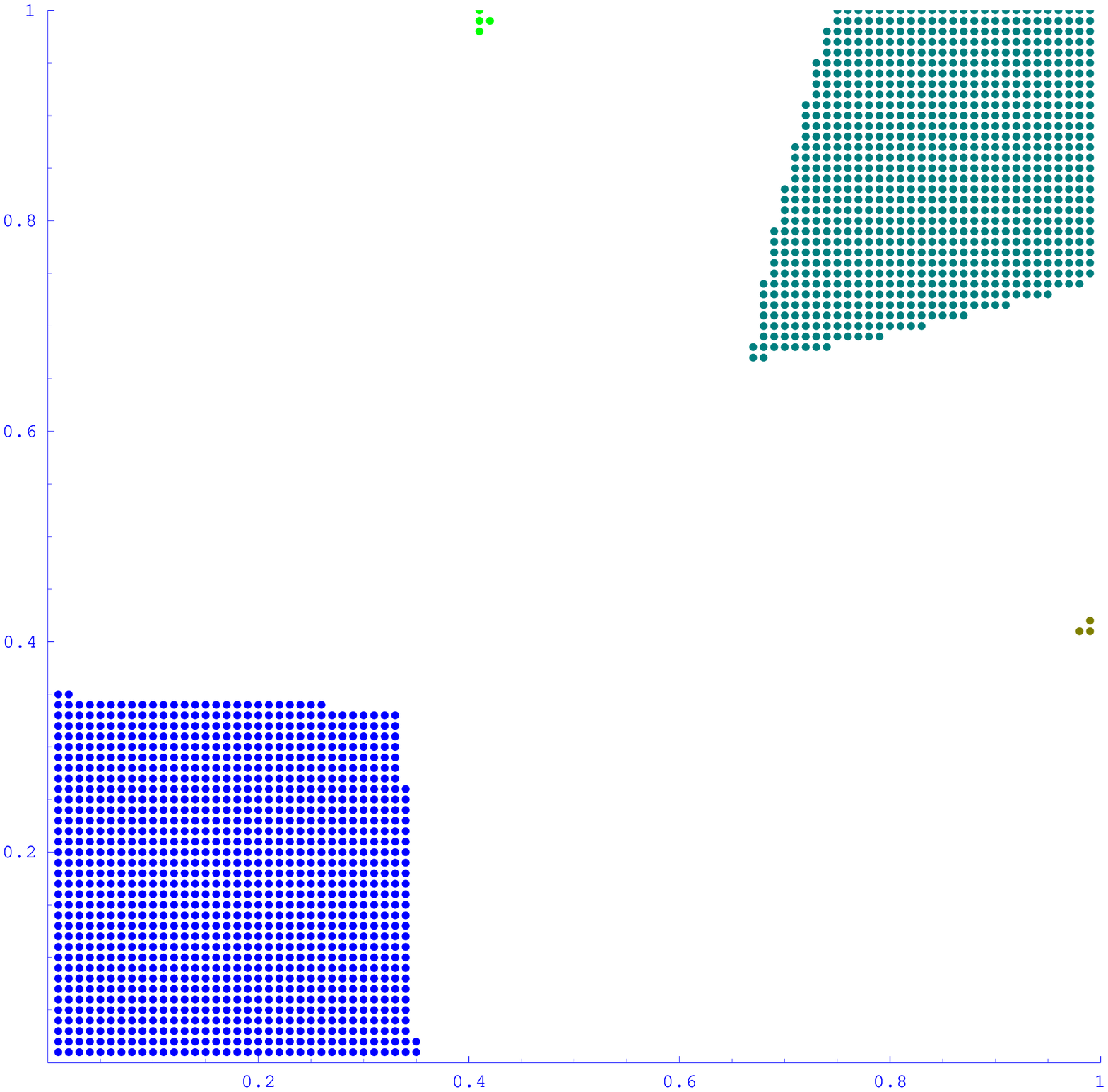}}
\endpspicture\cr
\tableCaption\cosTh{map of the stability zones at energy $E=-.5$ in the square 
$[0,1]^2$ sampled at a resolution $N=100$. At this energy just the four zones
shown are left.}\cr
}}
$$

\vfill\eject

\null
\vskip 1.cm
$$
\vbox{\halign{\hfill#\hfill\cr
\pspicture(0,0)(16,16)
\rput(8,8){\epsfxsize=15cm\epsfbox{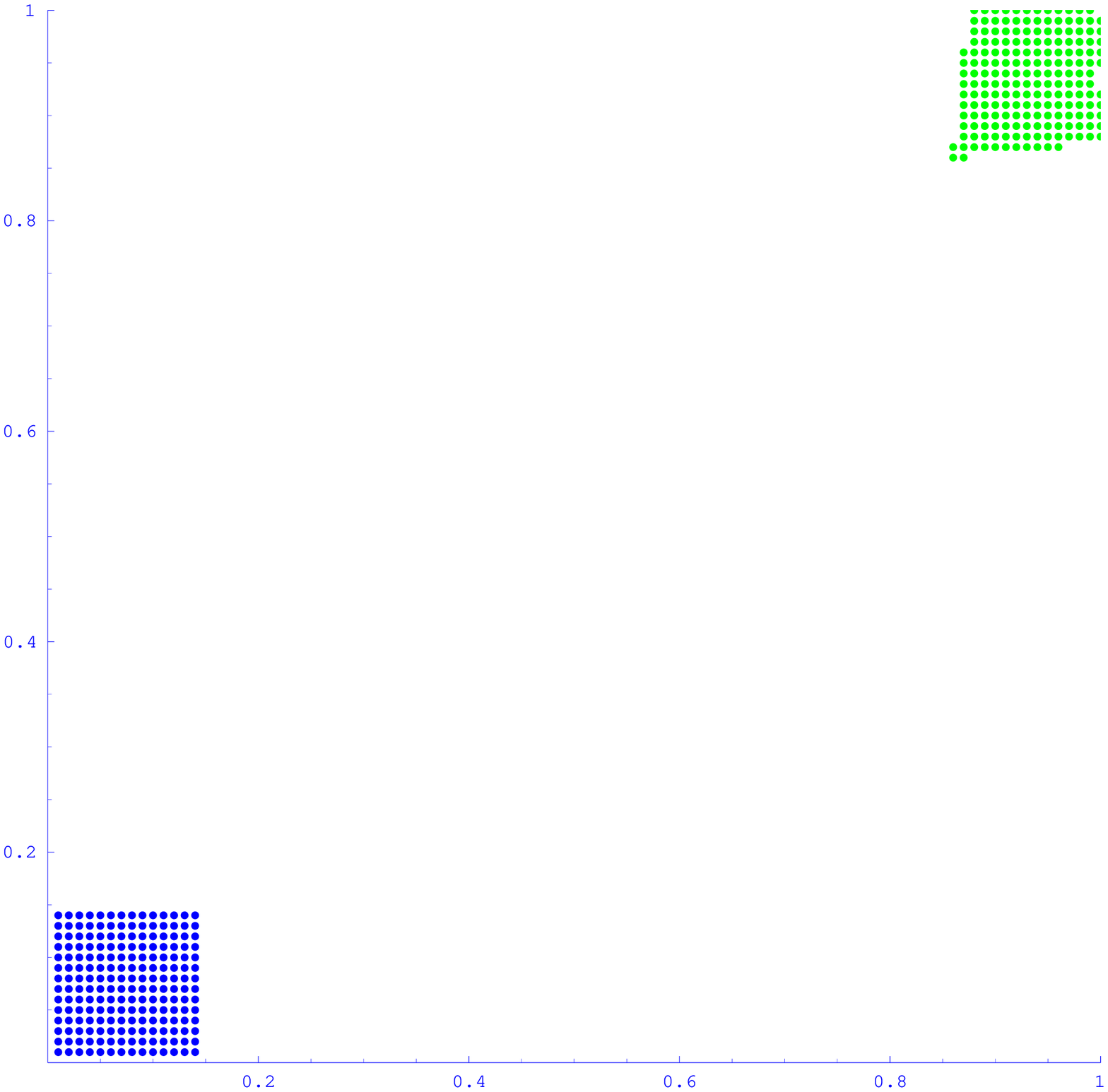}}
\endpspicture\cr
\tableCaption\cosTh{map of the stability zones at energy $E=-.9$ in the square 
$[0,1]^2$ sampled at a resolution $N=100$. Just the two biggest zones are now
visible.}\cr
}}
$$

\vfill\eject

\null
$$
\vbox{\halign{\hfill#\hfill\cr
\pspicture(0,0)(16,16)
\rput(8.5,8.5){\epsfxsize=20.5cm\epsfbox{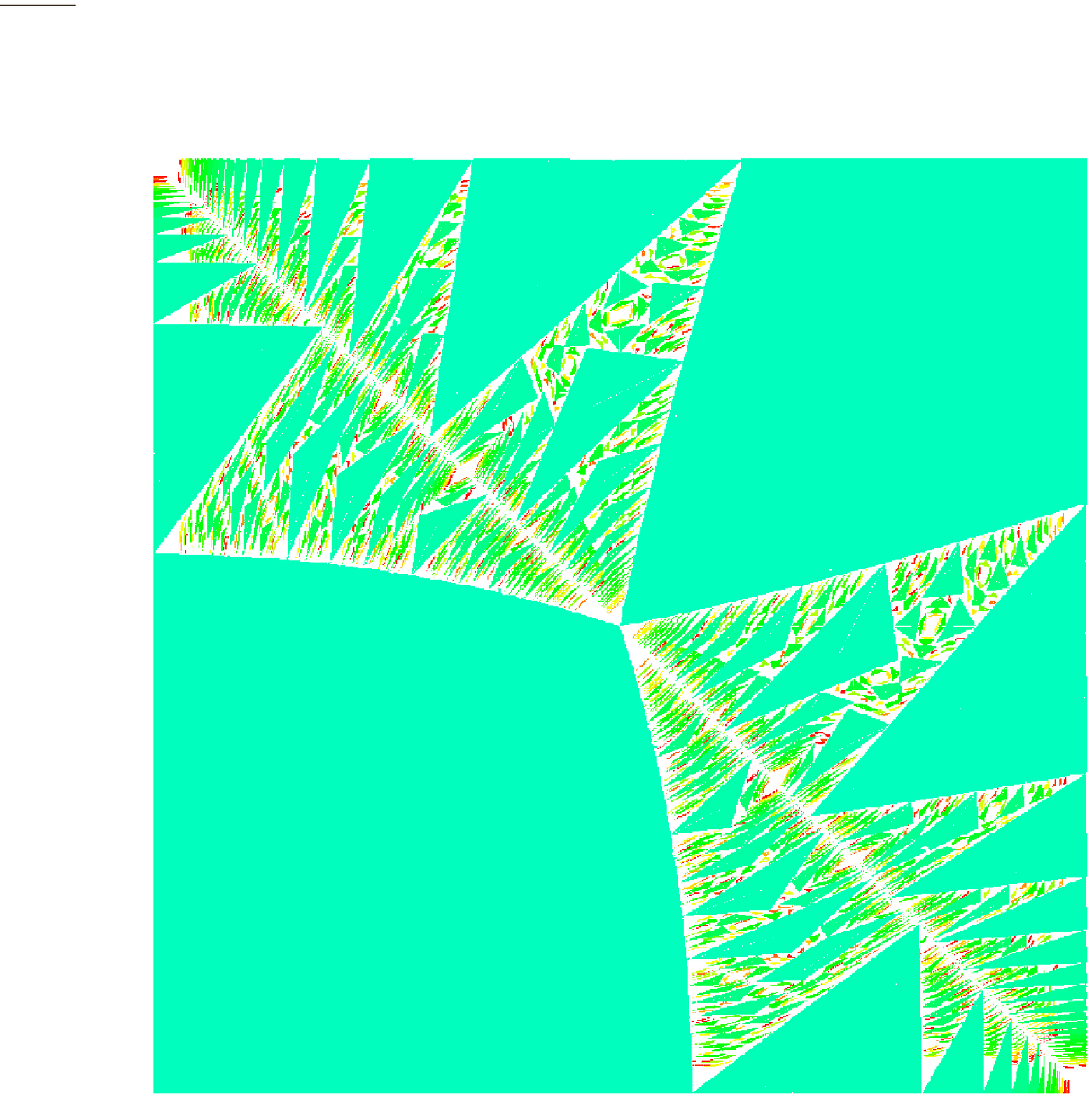}}
\endpspicture\cr
\tableCaption\perParFractal{the fractal picture for the piecewise quadratic 
function in the square $[0,1]^2$ obtained at a resolution $N=10^3$. 
Of the $\sim3\cdot10^4$ zones found just the
ones with at least 10 points ($\sim1000$) are shown. The square has been 
obtained just symmetrizing the triangular picture obtained.}\cr
}}
$$

\vfill\eject

\null

$$
\vbox{\halign{\hfill#\hfill\cr
\pspicture(0,0)(16,16)
\rput(8,8){\epsfxsize=15cm\epsfbox{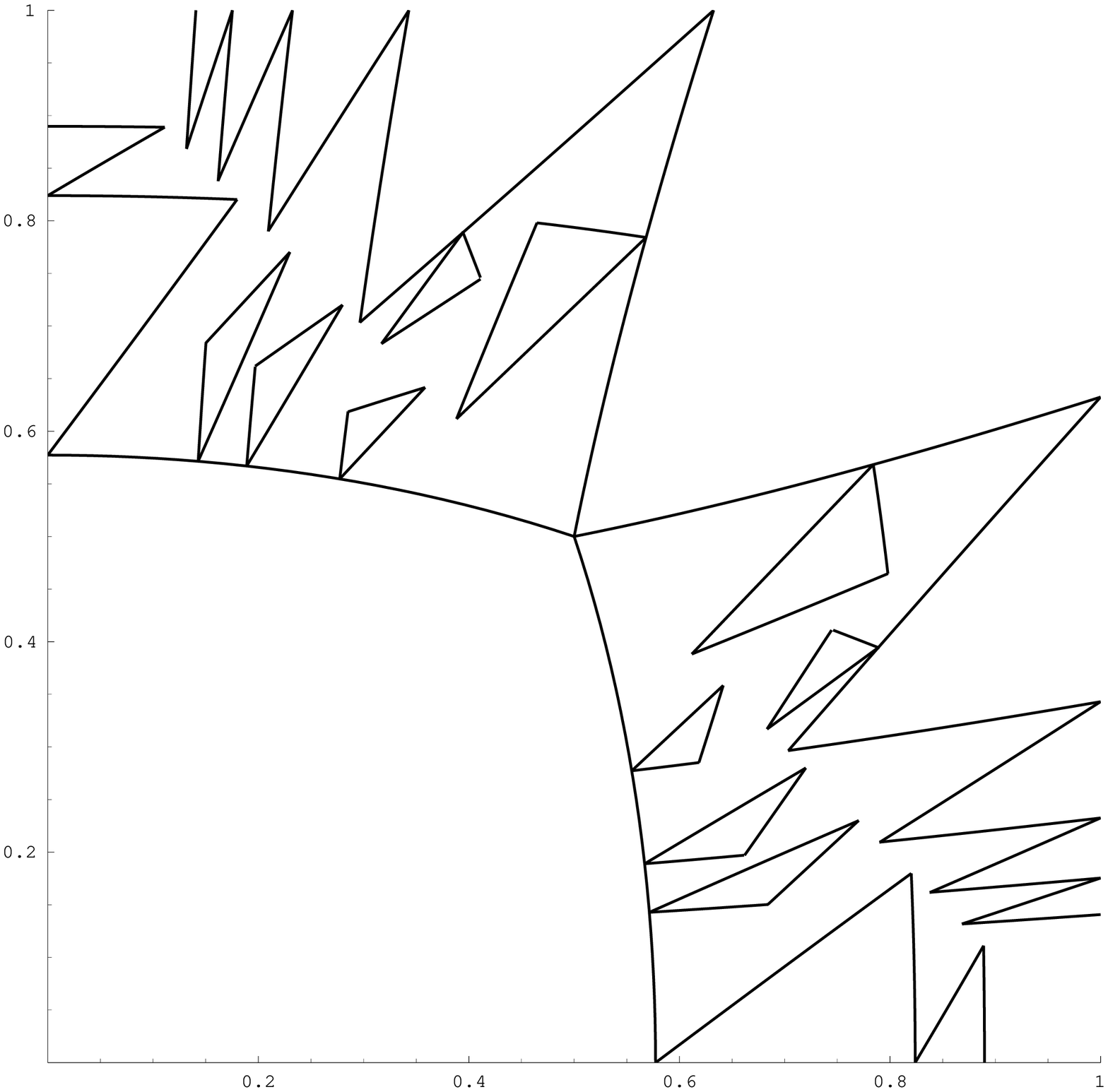}}
\rput(4.93,4.83){\rm(0,0,1)}
\rput(12.22,12.12){\rm(1,1,1)}
\rput(6.96,14.09){\rm(1,2,2)}
\rput(14.19,6.86){\rm(2,1,2)}
\rput(1.9,11.46){\rm(0,1,2)}  
\rput(12,1.9){\rm(1,0,2)}
\rput(4.9,14.55){\sixrm(1,3,3)}
\rput(14.65,4.9){\sixrm(3,1,3)}
\rput(11.6,7.67){\sixrm(3,2,4)}
\rput(7.84,11.49){\sixrm(2,3,4)}
\rput(3.55,10.7){\fiverm(1,4,6)}
\rput(4.2,10.3){\sixrm(1,3,5)}
\rput(10.15,4){\sixrm(3,1,5)}
\rput(10.69,3.4){\sixrm(4,1,6)}
\rput(1.7,13.5){\sixrm(0,2,3)}
\rput(13.45,1.28){\sixrm(2,0,3)}
\rput(4.,15.1){\sixrm(1,4,4)}
\rput(6.65,11.7){\fiverm(2,4,5)}
\rput(11.9,6.55){\sixrm(4,2,5)}
\rput(14.84,3.7){\sixrm(4,1,4)}
\rput(3.35,15.4){\sixrm(1,5,5)}
\rput(14.95,3.15){\sixrm(5,1,5)}
\rput(5.47,9.7){\fiverm(1,2,4)}
\rput(9.7,5.2){\fiverm(2,1,4)}
\endpspicture\cr
\tableCaption\parSk{picture analogous to the one in table \cosSk\ in case
of the piecewise quadratic function.}\cr 
}}
$$

\vfill\eject

\null
\vskip 1.cm
$$
\vbox{\halign{\hfill#\hfill\cr
\pspicture(0,0)(16,16)
\rput(8,8){\epsfxsize=15cm\epsfbox{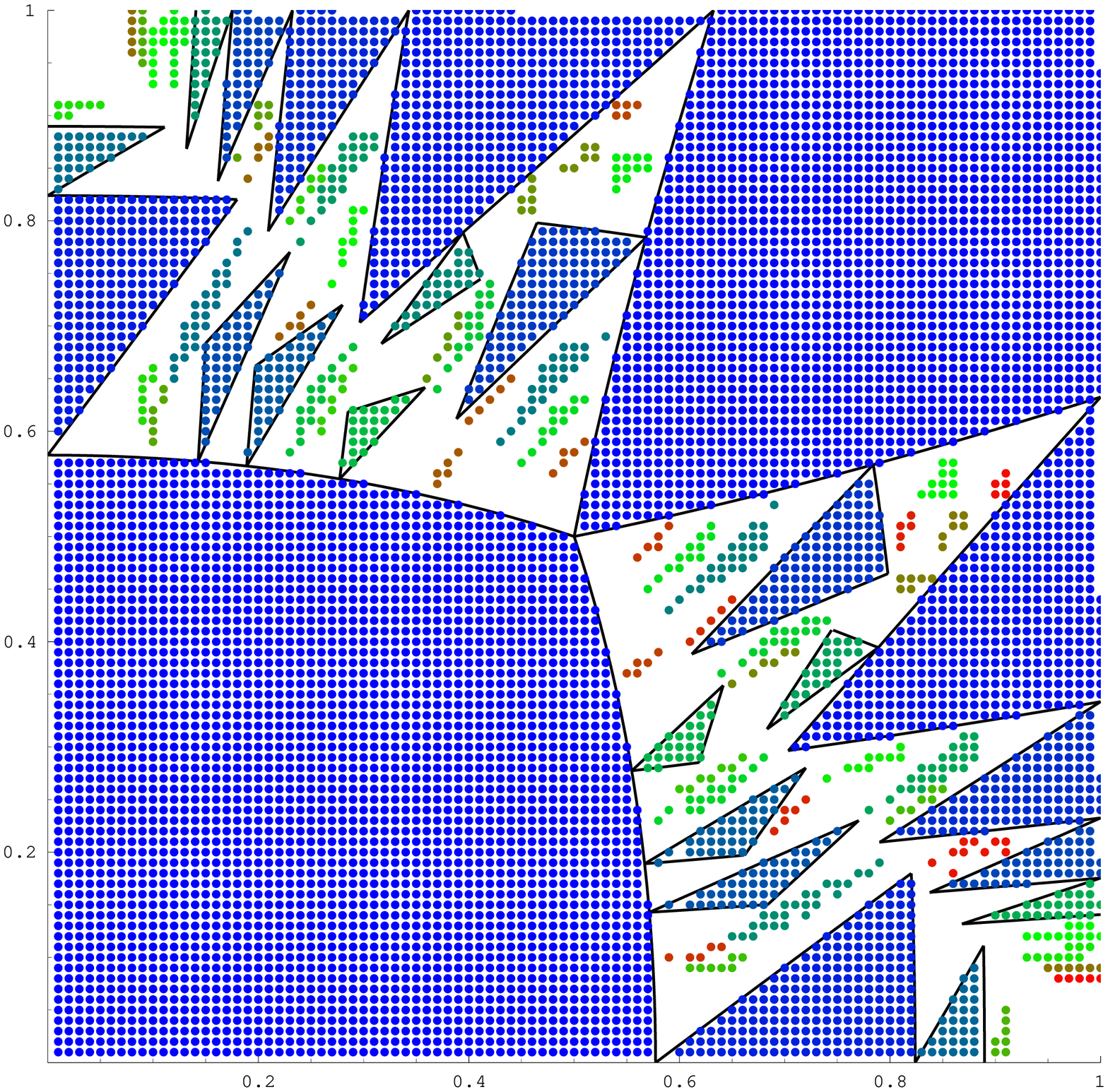}}
\endpspicture\cr
\tableCaption\primitiveCell{map of the stability zones for the piecewise
quadratic function in the square $[0,1]^2$ sampled at $E=0$ with resolution $N=100$. 
Of the 708 zones found, just the 74 with more than 5 points are shown. 
The boundary found analytically as explained in section \cosTd is also 
shown for a few zones to show the perfect agreement with the numerical results.
They are very close to the boundaries of trigonometric function shown in
table \cosSk\ and the homology zones that labels them are exactly the same 
than in the trgonometric case.}\cr
}}
$$

\vfill\eject

\null
\vskip 1.cm
$$
\vbox{\halign{\hfill#\hfill\cr
\pspicture(0,0)(16,16)
\rput(8,8){\epsfxsize=15cm\epsfbox{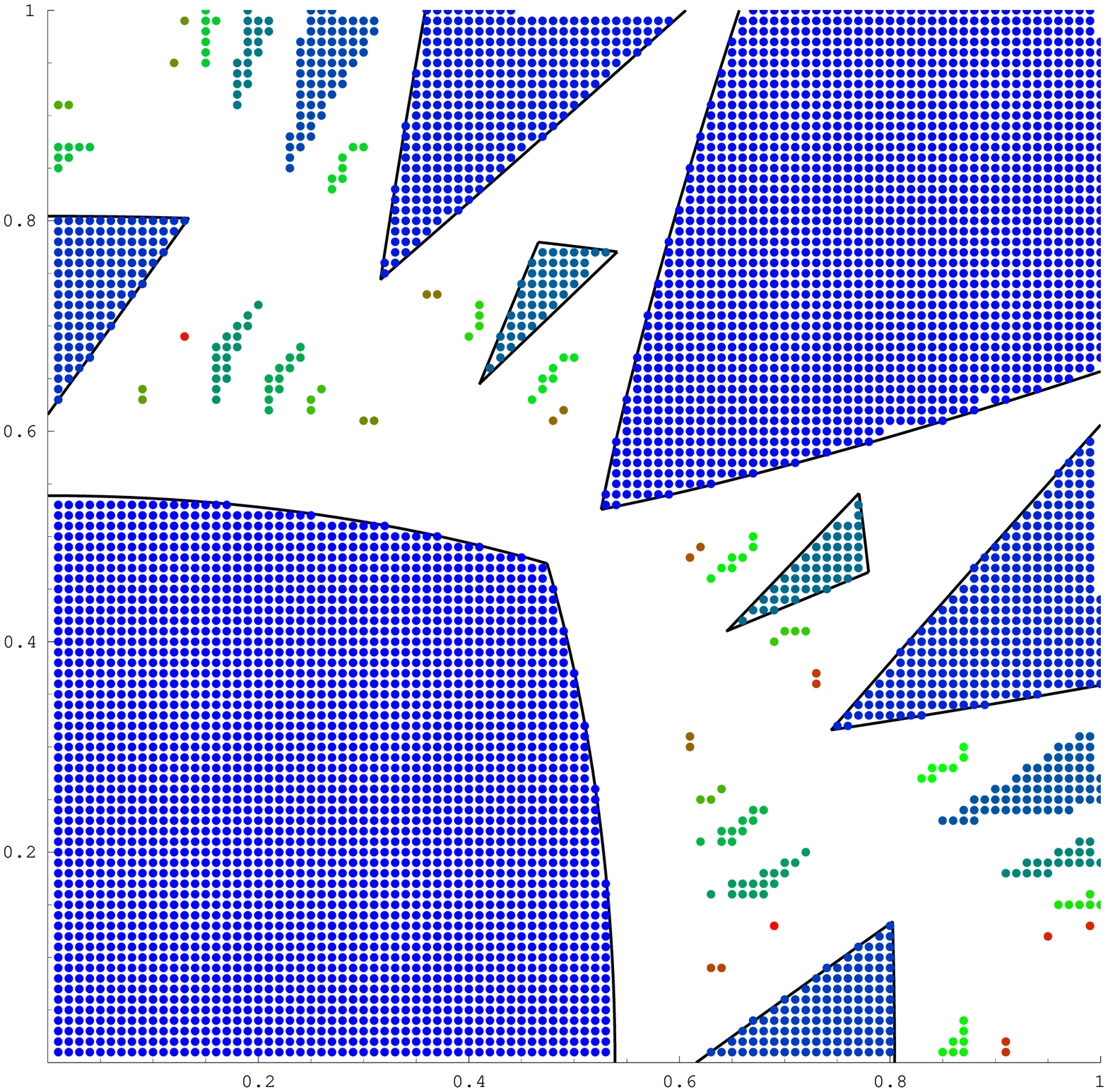}}
\endpspicture\cr
\tableCaption\primitiveCell{map of the stability zones for the piecewise
quadratic function in the square $[0,1]^2$ sampled at $E=-.1$ with resolution $N=100$. 
We show here all 42 zones found together with the boundaries of the
biggest ones.}\cr
}}
$$

\vfill\eject

\null
\vskip 1.cm
$$
\vbox{\halign{\hfill#\hfill\cr
\pspicture(0,0)(16,16)
\rput(8,8){\epsfxsize=15cm\epsfbox{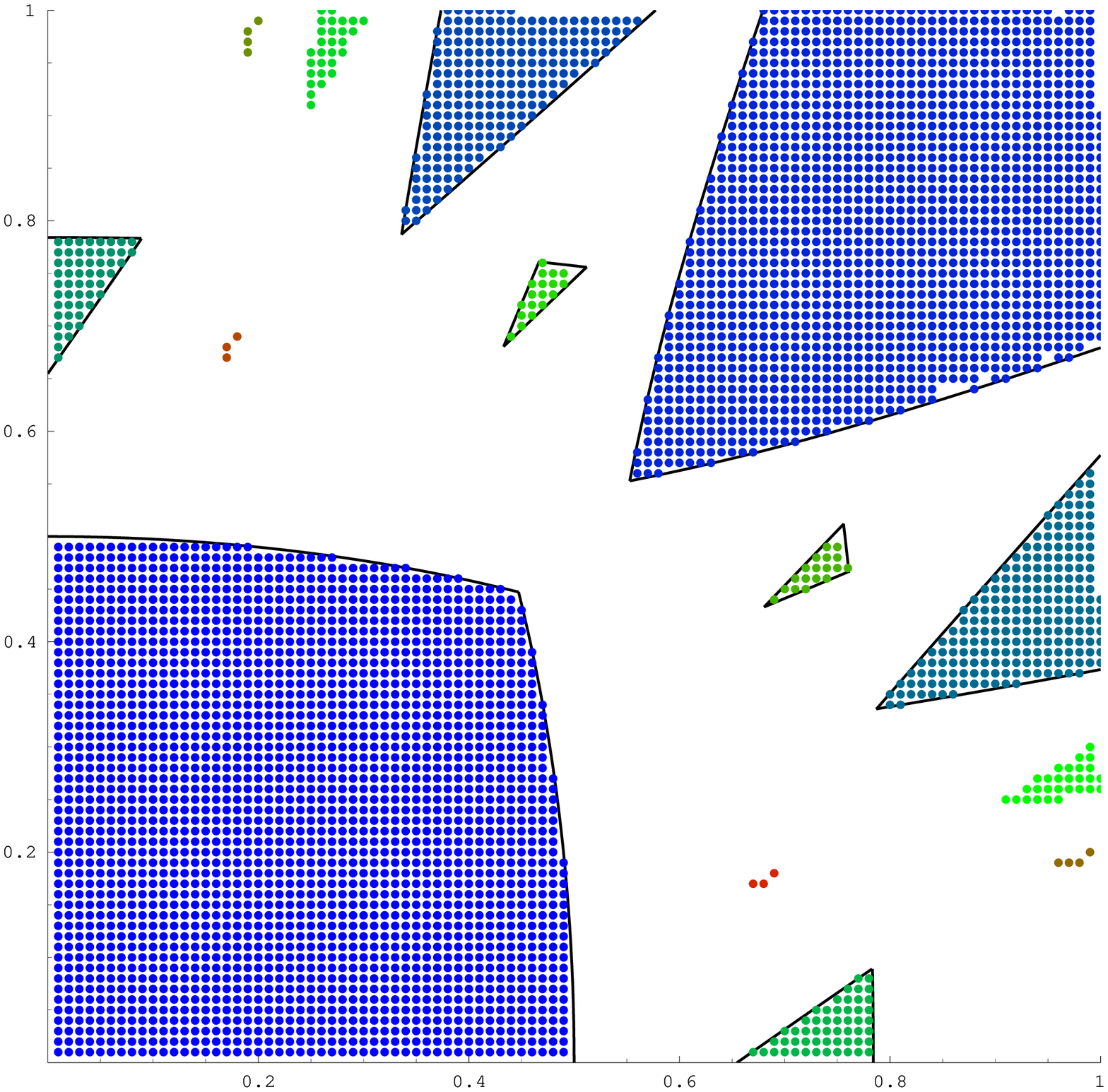}}
\endpspicture\cr
\tableCaption\primitiveCell{map of the stability zones for the piecewise
quadratic function in the square $[0,1]^2$ sampled at $E=-.2$ with resolution $N=100$. 
All 14 zones found are shown together with boundaries of the biggest ones.}\cr
}}
$$

\vfill\eject

\null
\vskip 1.cm
$$
\vbox{\halign{\hfill#\hfill\cr
\pspicture(0,0)(16,16)
\rput(8,8){\epsfxsize=15cm\epsfbox{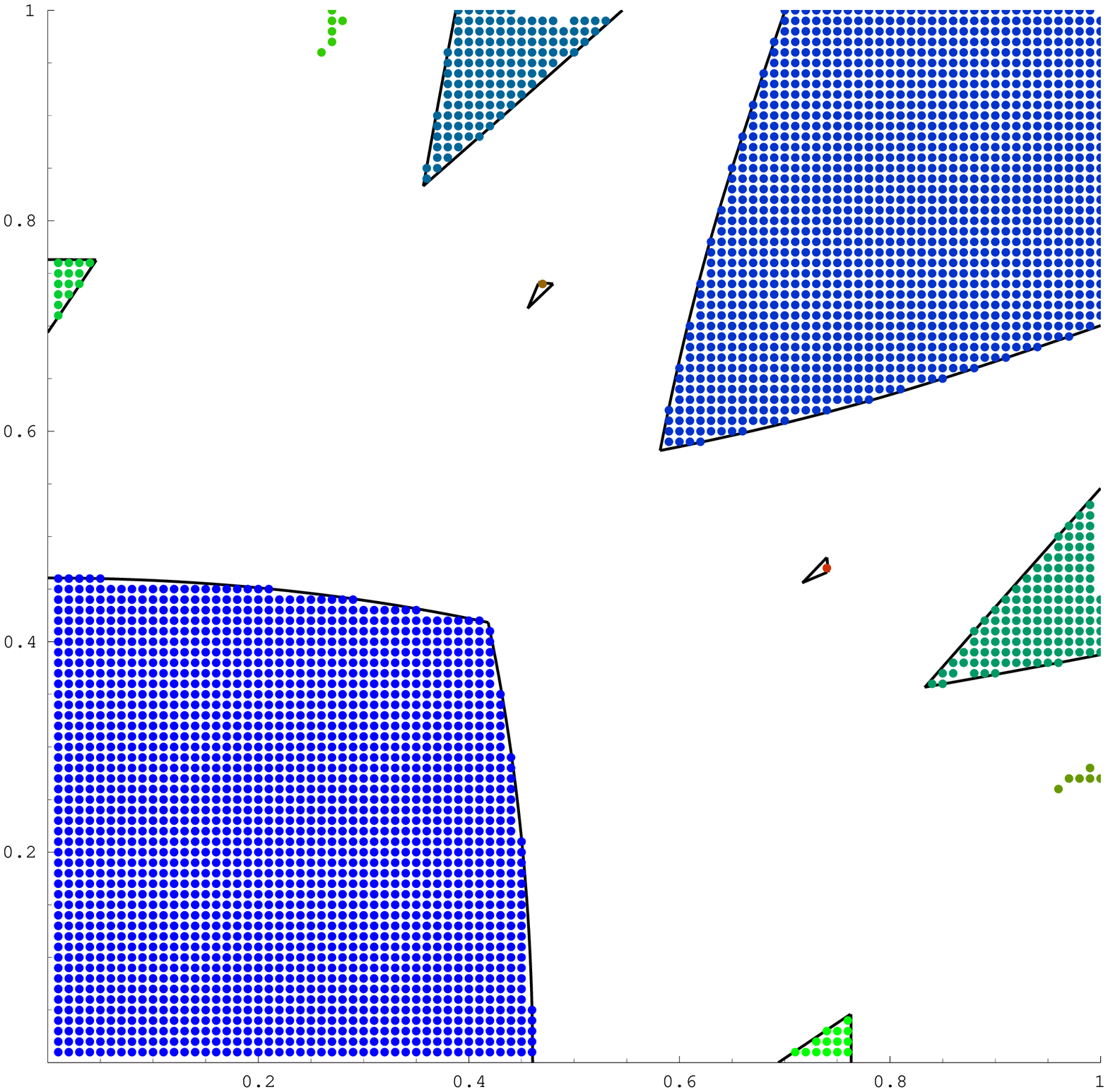}}
\endpspicture\cr
\tableCaption\primitiveCell{map of the stability zones for the piecewise
quadratic function in the square $[0,1]^2$ sampled at $E=-.3$ with resolution $N=100$. 
All 10 zones found are shown together with boundaries of the biggest ones.}\cr
}}
$$

\vfill\eject

\null
\vskip 1.cm
$$
\vbox{\halign{\hfill#\hfill\cr
\pspicture(0,0)(16,16)
\rput(8,8){\epsfxsize=15cm\epsfbox{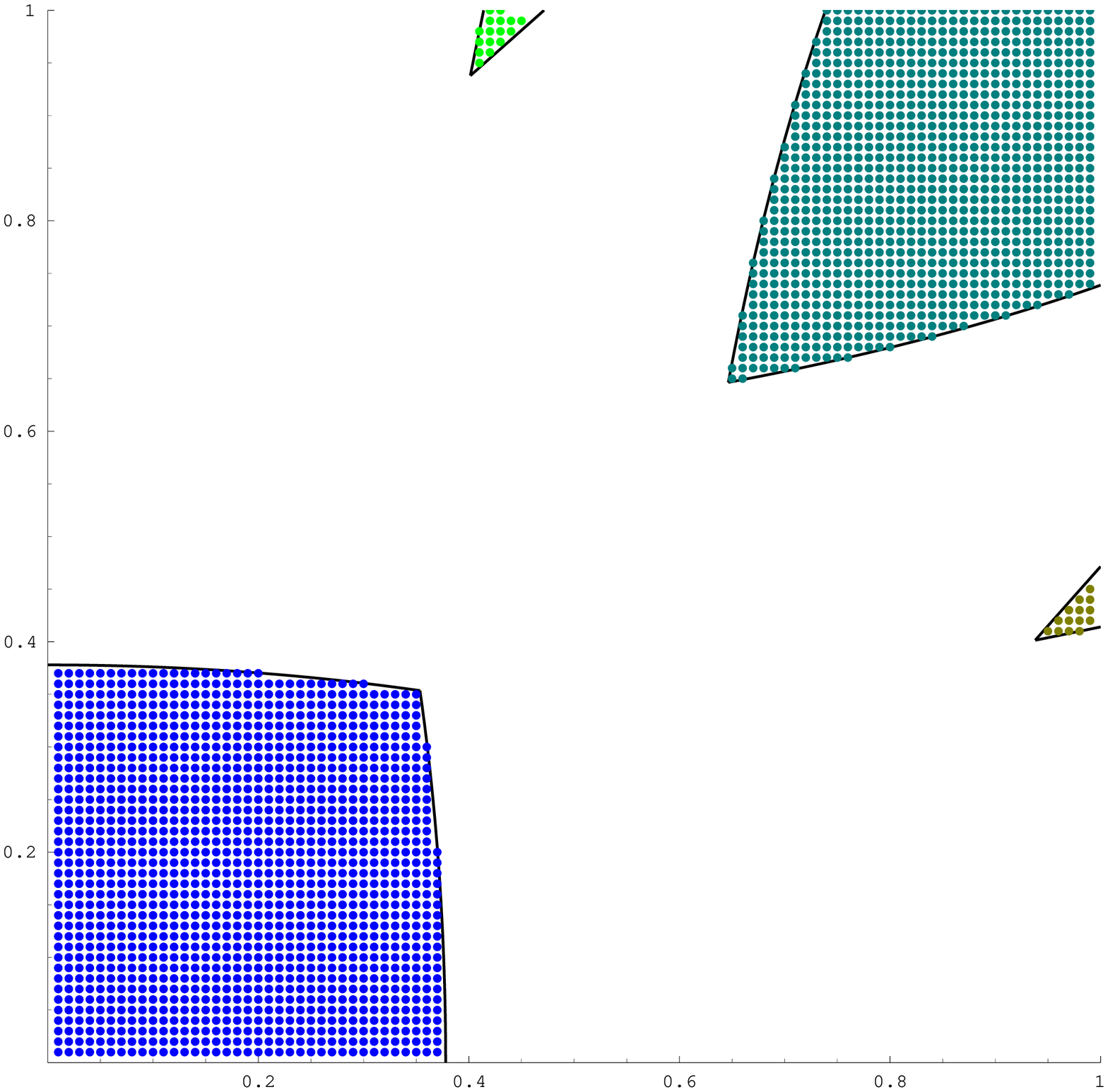}}
\endpspicture\cr
\tableCaption\primitiveCell{map of the stability zones for the piecewise
quadratic function in the square $[0,1]^2$ sampled at $E=-.5$ with resolution $N=100$. 
All 4 zones found are shown together with their boundaries.}\cr
}}
$$

\vfill\eject

\null
\vskip 1.cm
$$
\vbox{\halign{\hfill#\hfill\cr
\pspicture(0,0)(16,16)
\rput(8,8){\epsfxsize=15cm\epsfbox{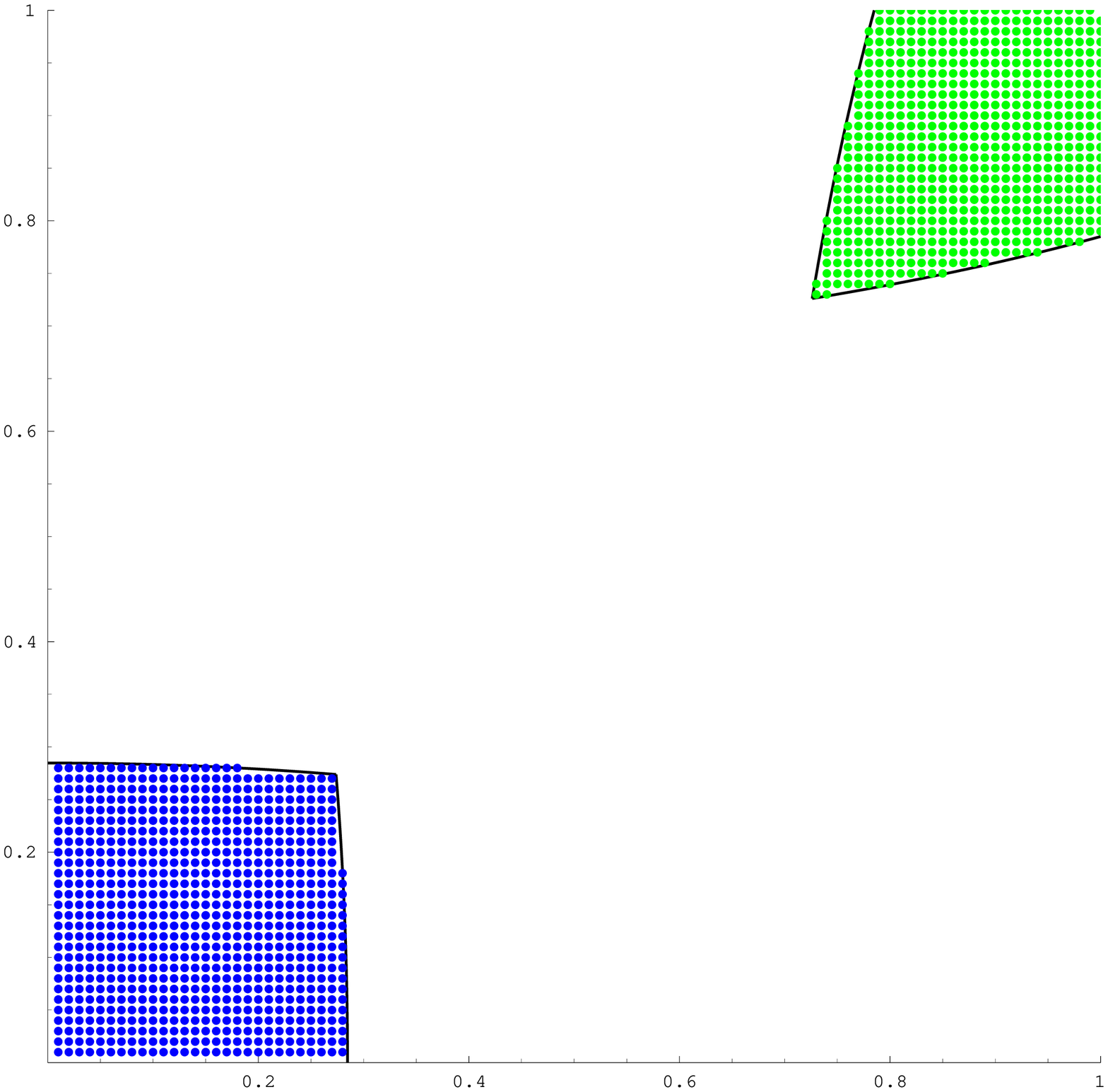}}
\endpspicture\cr
\tableCaption\primitiveCell{map of the stability zones for the piecewise
quadratic function in the square $[0,1]^2$ sampled at $E=-.7$ with resolution $N=100$. 
Just the two main zones survive at this energy.}\cr
}}
$$

\vfill\eject

\null
\vskip 1.cm
$$
\vbox{\halign{\hfill#\hfill\cr
\pspicture(0,0)(16,16)
\rput(8,8){\epsfxsize=15cm\epsfbox{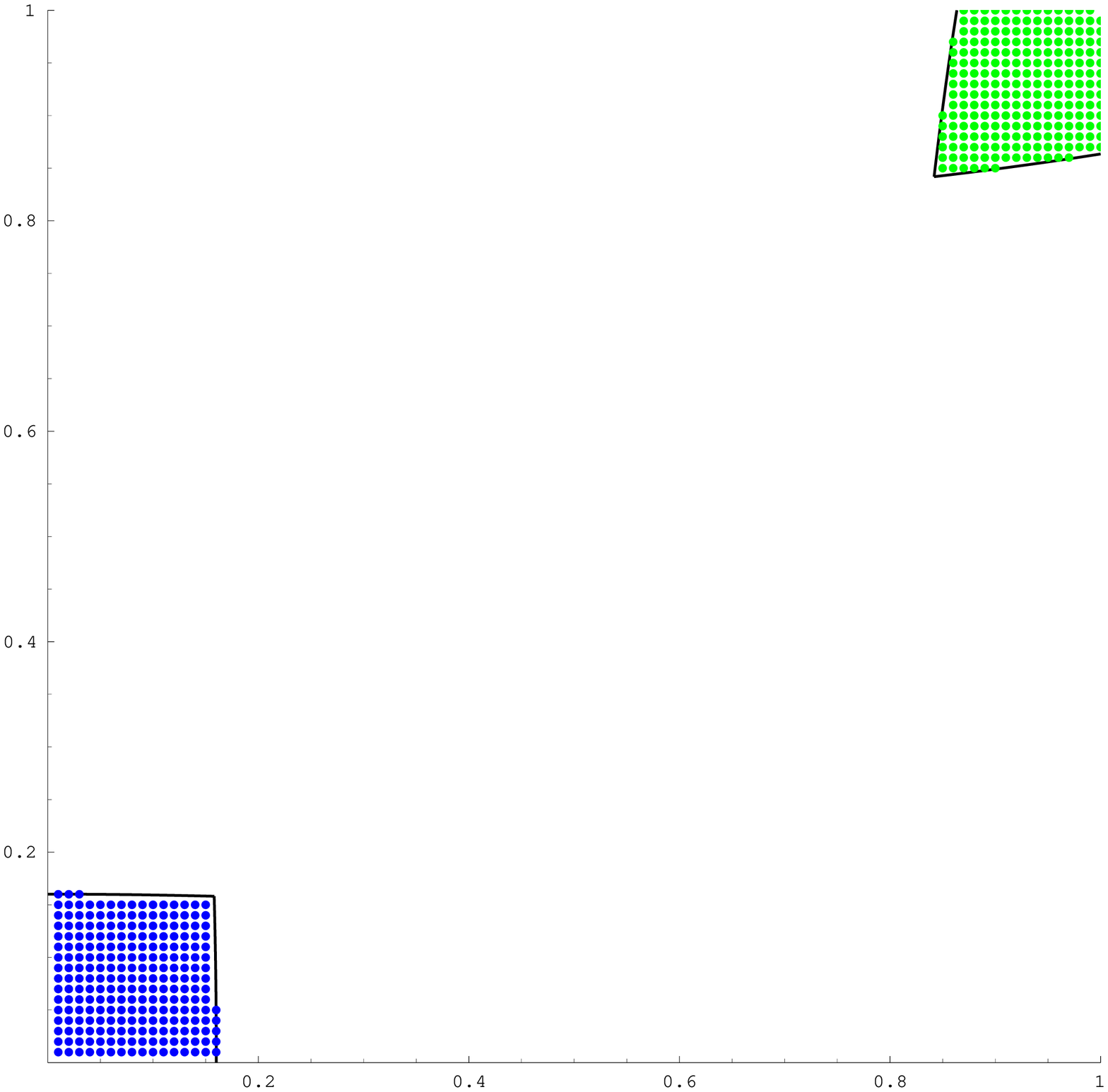}}
\endpspicture\cr
\tableCaption\primitiveCell{map of the stability zones for the piecewise
quadratic function in the square $[0,1]^2$ sampled at $E=-.9$ with resolution $N=100$. 
Just the two main zones survive at this energy.}\cr
}}
$$

\vfill\eject\end